\documentstyle[aps,prb,preprint,epsf]{revtex}
\newcommand{\leqslant}{\leq}
\newcommand{\geqslant}{\geq}
\begin{document}

\draft
\title{Dynamic Critical Phenomena in Channel Flow}
\author{Joe Watson and Daniel S. Fisher}
\address{Department of Physics, Harvard University, 
         Cambridge, Massachusetts 02138}
\date{Draft of \today}
\maketitle

\begin{abstract}
  A simple model of the driven motion of interacting particles in a
  two dimensional random medium is analyzed, focusing on the critical
  behavior near to the threshold that separates a static phase from a
  flowing phase with a steady-state current. The critical behavior is
  found to be surprisingly robust, being independent of whether the
  driving force is increased suddenly or adiabatically. Just above
  threshold, the flow is concentrated on a sparse network of channels,
  but the time scale for convergence to this fixed network diverges
  with a larger exponent that that for convergence of the current
  density to its steady-state value. This is argued to be caused by
  the ``dangerous irrelevance'' of dynamic particle collisions at the
  critical point. Possible applications to vortex motion near to the
  critical current in dirty thin film superconductors are discussed
  briefly.
\end{abstract}

%%%%%%%%%%%%%%%%%%%%%%%%%%%%%%%%%%%%%%%%%%%%%%%%%%%%%%%%%%%%%%%%%%%
\section{Introduction}
\label{sec:intro}

Vortices in superconductors that are driven by an applied electric
current and pinned by disorder exhibit a wide range of interesting
behavior. These are examples of non-linear transport in random media
which is collective in that the interactions between the transported
can roughly be divided into two classes: ``elastic'' and ``plastic''
depending on whether or not the interactions between the transported
particles are strong enough to maintain the particles in an extended
elastic structure as they move.  Our interest here is plastic flow
where the interactions between the particles and the random medium
(``pinning'') are strong enough to break up any elastic structure.  We
will focus on the very strong pinning limit for which ``channel flow''
can occur: where the flow is not only plastic but dominated by
particles moving along a sparse network of persistent channels.

Plastic flow of vortices has attracted a lot of recent interest%
.\cite{nori:scirev,matsuda:sciriv} %
A number of experimental measurements in $2\mbox{H}$-$\mbox{NbSe}_2$
have been attributed to plastic flow, including an unexpected dip in
the electrical resistance just below $H_{c2}$ (``peak effect'')%
,\cite{bhatt:peakeffect,higgins:vardyn} %
anomalous $I$-$V$ curves%
,\cite{bhatt:dyndis} %
generation of unusual broadband noise%
,\cite{marley:ffnoise} %
small angle neutron scattering measurements%
,\cite{yaron:twostep+yaron:twostepadd} %
and fingerprint phenomena where the detailed shape of the $I$-$V$
curve is repeatable for a single sample but differs between samples%
.\cite{bhatt:fingerprint} %
Similar phenomena have recently been observed in two-dimensional
amorphous $\mbox{Mo}_{77}\mbox{Ge}_{23}$ films %
\cite{hellerqvist:vdynmoge} %
which are supported by
numerical simulations for two-dimensional systems%
\cite{%
gronbech:filamoge,%
jensen:plaslett,%
jensen:plaslong,%
shi:pininvfl,%
faleski:vortdyn} %
which clearly see vortex motion dominated by flow along narrow
``filamentary'' channels.  In addition to this indirect evidence,
realtime images of moving vortices in thin films have been
recorded%
.\cite{%
harad:rtvortex,%         92 nature
harad:vortexmicroscopy,% 93 prl
harad:rtjjap,%           94 jjap
matsuda:sciriv} %        96 science
which clearly show individual vortices moving along narrow paths
of least resistance.

Some of these experiments also suggest that in the regimes studied
plastic flow only persists for forces just above the threshold for the
onset of motion, i.e., near the critical current. At higher current
the vortex lattice appears to become more ordered. Previous
theoretical work%
\cite{koshelev:dynmelt,%
balents:tempord} %
that studied plastic flow has primarily concentrated on the ordering
and break up of an elastic lattice by the randomness, concluding that,
at least for small driving forces in two dimensions, the lattice will
always break up and the flow will become plastic.  Here we take a
complementary approach that considers the extreme limit of a fully
plastic state with only hard-core intervortex interactions.

In this paper we extend work on a simple model%
\cite{narayan:nlflow,watson:disc} %
---roughly of point vortices in a thin film.  It exhibits two phases:
if the driving force is small the vortices are all trapped and there
is no steady-state current, but if the force exceeds a finite
threshold the vortices move in a static channel network whose
configuration is determined by the pinning in the sample. In the
steady-state just above threshold, the channels are far apart but each
channel carries a high vortex current density.

We focus on the dynamic critical behavior near threshold and on the
development of the channel network above threshold. This study will be
primarily numerical aided by scaling analysis. The critical behavior
appears to represent a universality class of non-equilibrium ``phase''
transitions that probably also includes some aspects of a continuum
fluid model studied by Narayan and Fisher%
.\cite{narayan:nlflow} %

%%%%%%%%%%%%%%%%%%%%%%%
\subsection{Outline}
\label{sec:outline}

In the rest of the Introduction we briefly explain the model system we
have investigated and then outline our main results.  In
Sec.~\ref{sec:modelimp} we provide a more detailed description of the
model. The dynamic critical phenomena and scaling properties are
explained in Sec.~\ref{sec:crit}, while the development of the channel
network is studied in Sec.~\ref{sec:conv}. Finally we compare our
results to related work and consider possible applications in
Sec.~\ref{sec:appconc}.

%%%%%%%%%%%%%%%%%%%%%%%
\subsection{Model}
\label{sec:modeintro}

The model studied here was introduced in Ref.~\onlinecite{watson:disc}
(where it was called the ``one-deep'' model). We briefly summarize its
features here and define it in more detail in Sec.~\ref{sec:modelimp}.

This simple model for the motion of vortices in a two-dimensional
random medium, treats the vortices as point particles with a hard-core
interaction which move on a lattice. The randomness is represented by
each site's capacity to trap arriving particles and by the existence
of a locally preferred direction---the ``primary outlet''---for motion
out of each site (see Fig.~\ref{fig:paths2}). The local capacity and
preferred direction can be thought of as barriers for a particle to
leave by one of the site's outlets. Particles only leave through a
particular outlet of a site when the number of particles present
exceeds the barrier height.  Increasing the driving force ($F$)
corresponds to lowering all of the barrier heights which causes some
sites to overflow thereby releasing particles. If one excess particle
is present on a site, it {\em always\/} leaves thought the primary
outlet. If more than one excess particle is present, one will leave
through each outlet---a ``split''. Decreasing the force raises
barriers, creating more traps and increasing the capacity of existing
traps.

Various possible histories of the system are represented by the way in
which the force is changed, in this paper we will mainly consider
increasing the force very quickly (``sudden'' forcing) and then
waiting for the particles to reach a steady-state.  But we will also
investigate what happens when the force is increased very slowly
(``adiabatic'' forcing) for comparison.

%%%%%%%%%%%%%%%%%%%%%%%%%%%%%%
\subsection{Dynamic critical behavior}
\label{sec:critintro}

The behavior of the system is divided into two phases: A flowing phase
where a steady-state current flows and a static phase with no current.
In the static phase, the system has two types of sites: saturated
sites at which any additional particle would overflow, and unsaturated
sites or ``traps''. In the static phase, the particles will respond to
a sudden increase in the force only in a transient manner, with some
fraction of the particles moving varying distances downhill before
being trapped at previously unsaturated sites. The static
configuration can be probed by adding a single particle, which, if it
is added to a saturated site, will flow through primary outlets down a
``drainage tree'' until it stops at the unsaturated site at the bottom
of the tree. At a critical force~$F_c$, the lengths of some of the
trees will diverge and above the critical force the system will have
steady-state flow.  A fundamental property of our model is that at
large times above threshold the particle flow converges to a fixed
channel network with each channel carrying its maximum capacity%
.\cite{watson:disc} %
This implies that there are three types of lattice sites in the
steady-state: network sites that carry moving particles, and
off-network sites with no particle flow which can be either in the
drainage basin (sites that drain to the channel network), or in finite
drainage trees.  The network sites form persistent channels that drain
the system. The channels can join and split as they move through the
system. In an infinite system, which sites are on the network (for a
particular current) depends {\em only\/} on which outlets are the
primary outlets and not on the initial particle placements.  

We now summarize the critical behavior. In the flowing phase in a
large system the steady-state current density (i.e., the mean number
of particles per lattice site), after a sudden increase in the force
to a final value of~$F$ just above $F_c$, is found to be
\begin{equation}
  J\sim (F-F_c)^{\beta}.
  \label{eq:currentJ}
\end{equation}

\begin{figure}
  \begin{center}
    \leavevmode
    \epsfxsize=4truein
    \epsfbox{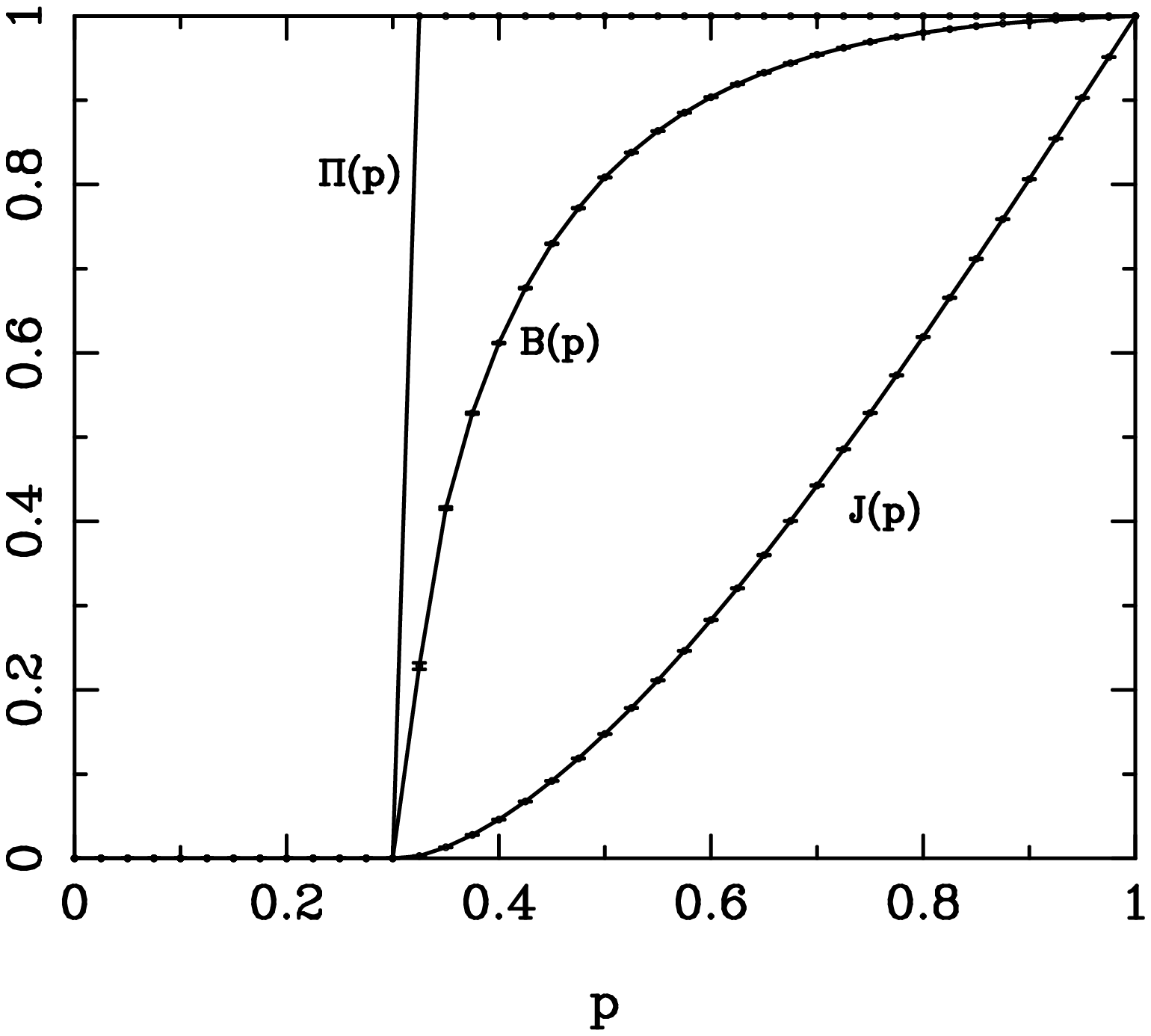}
  \end{center}
  \caption{Fraction of systems with a non-zero
  steady-steady current [$\Pi(p)$], steady-state current [$J(p)$] and
  fraction of sites in the drainage basin [$B(p)$] for a 1024x512
  lattice averaged over 32 systems.}
  \label{fig:all3}
\end{figure}

Another quantity of interest in the flowing phase is the fraction of
sites in the drainage basin, i.e., sites to which an added particle,
following primary outlets, would continue to move indefinitely and
become part of the steady-state current. The fraction of sites in the
basin is found to increase as
\begin{equation}
  \label{eq:pinftperc}
  B\sim (F-F_c)^{\Gamma}.
\end{equation}
Typical results from a set of simulations showing this behavior are
shown in Fig.~\ref{fig:all3}.

Numerical simulations of the model yield values of these exponents
\begin{eqnarray}
  \beta  &=& 1.53\pm0.03,\\
  \Gamma &=& 0.49\pm0.02.
\end{eqnarray}
These error bars, and all uncertainties quoted and plotted here, are
estimates of statistical 1-$\sigma$ errors. More details of the
accuracy and independence of these measurements are given in
Sec.~\ref{sec:crit}. 

%%%%%%%%%%%%%%%%%%%%%%%%%%%%%%
\subsubsection{Length scales}

We are, of course, primarily interested in the behavior of infinite
systems. But numerical studies of a range of system sizes yield {\em
  both\/} ways to extrapolate to infinite systems and information on
the characteristic length scales in the system.  Our scaling is
anisotropic in the two directions, we will generally measure length
($L$) in the downhill direction and widths ($W$) in the horizontal
direction.

The behavior of a finite system of length~$L$ is controlled by a
correlation length $\xi_f(F)$ as described in Sec.~\ref{sec:crit}.
This length scale controls the finite size scaling and is found to
diverge with an exponent~$\nu$ as $F\rightarrow F_c$. In order to
obtain the critical force and the correlation length it is convenient
to work with $\Pi(L,W,F)$ defined as the probability that there is a
steady-state current for a system of length~$L$ and width~$W$ at
force~$F$.  Using the scaling of $\Pi$ we find
\begin{equation}
  \label{nu}
   \nu=1.62\pm0.04
\end{equation}
on both sides of the threshold. 

\begin{figure}
  \begin{center}
    \leavevmode
    \epsfxsize=4truein
    \epsfbox{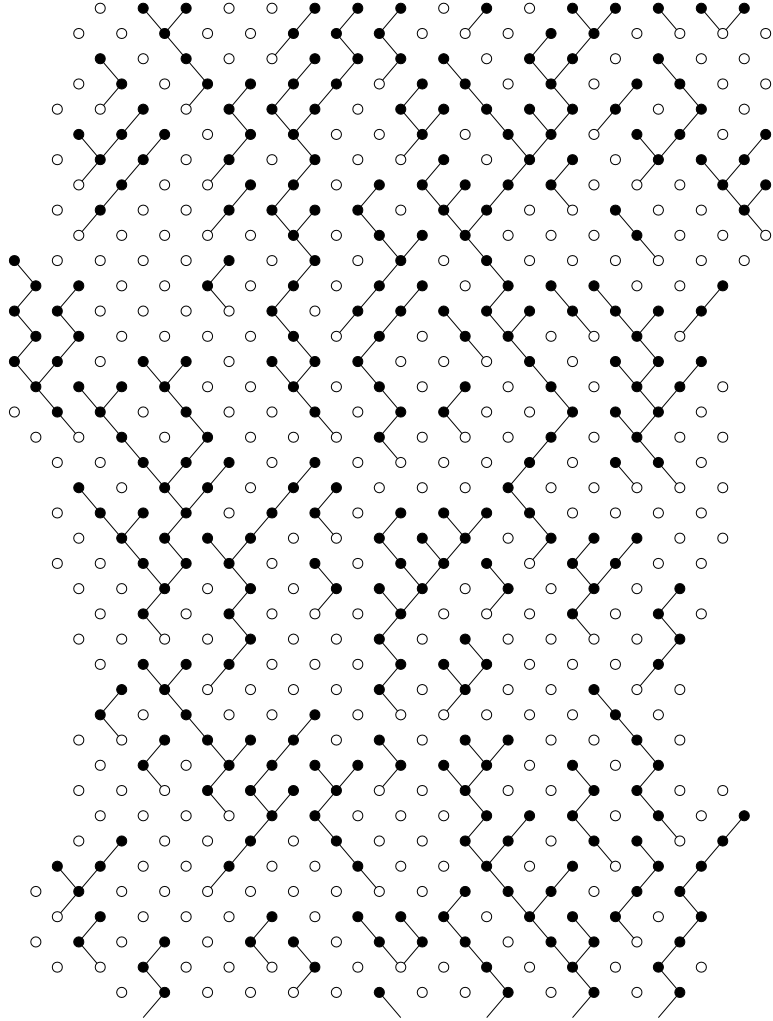}
  \end{center}
  \caption{Lattice model below threshold.
  The black sites are saturated and the white sites are unsaturated.
  The lines show the primary outlets of each saturated site and form
  drainage trees. The force direction is down the page.}
  \label{fig:latbelow}
\end{figure}

Below threshold at long times all particles are at rest and the sites
can be designated saturated and unsaturated (``traps'') depending on
whether an added particle will flow out of the site.  The saturated
sites and their primary outlets (see Fig.~\ref{fig:latbelow}) form
connected drainage trees such that an excess particle added to a site
on a drainage tree comes to rest on the unsaturated terminus site at
the bottom of the tree. The only important length scale is the
characteristic tree length above which trees are exponentially
unlikely, this we identify as proportional to~$\xi_f$.

However, above threshold the situation is rather more complicated:
there are {\em two\/} length scales present in the channel network.
The characteristic vertical distance scale of the channel network
backbone, $\xi_{n}$, which, as shown in Ref.~\onlinecite{watson:disc},
diverges as
\begin{equation}
    \xi_{n}\sim\frac{1}{J^2}
\end{equation}
as $J\rightarrow0$; and the length which controls finite size scaling
($\xi_f$).  If $\xi_n$ was identified with $\xi_f$ then we would
expect $\nu=2\beta$ which is clearly ruled our by the data; instead we
find $\beta/\nu=0.96\pm0.02$. Nevertheless, $\xi_n$ definitely
controls various physical quantities which suggests that both lengths
are important above threshold.

The presence of two different correlation lengths above threshold is
rather unusual (although it has appeared in other non-equilibrium
dynamics critical points, for example sliding charge density waves%
.\cite{narayan:scdweps}) %
In our system the two lengths control two different equilibration
processes.  The decay of current transients is rapid with
characteristic time~$\xi_f$, however redistribution of the current to
the steady-state channel network is much slower.  The convergence to a
fixed channel network can be seen by studying the distribution of
time-averaged local currents through different sites of the lattice.
For large systems at long times most of the sites either {\em
  always\/} contain moving particles or {\em never\/} contain moving
particles. This distribution has been studied numerically and is is
indeed found to converge to two delta functions (at currents of zero
and one) as the system size increases with characteristic time scale
of order $\xi_n$ and a power law tail ($1/t^{1/4}$) at long times.
Because this approach to a steady-state pattern is so slow near
threshold (and $\xi_{n}\gg\xi_{f}$) experiments or simulations may not
reach the steady-state of the current network even if the current
itself does reach its steady-state value.

The forcing in our model is strongly anisotropic and particles move
much further parallel to the force (downhill) than perpendicular to
it. The correlation lengths~$\xi_f$ and~$\xi_{n}$ are defined parallel
to the force. For perpendicular (horizontal) distances we define two
characteristic lengths~$\xi_{f_\perp}$ and~$\xi_{n_\perp}$ which we
expect to behave as
\begin{equation}
  \xi_{f_\perp}\sim\xi_f^{\alpha}
\end{equation}
and
\begin{equation}
  \xi_{n_\perp}\sim\xi_{n}^{\alpha_n}.
\end{equation}
From Ref.~\onlinecite{watson:disc}, the properties of the channel
network fix
\begin{equation}
  \alpha_n=\frac{1}{2}.
\end{equation}
Numerically, we obtain
\begin{equation}
  \alpha=0.50\pm0.05.
\end{equation}

%%%%%%%%%%%%%%%%%%%%%%%%%%%%%%%%
\subsubsection{Adiabatic forcing and universality class}

Our model seems to represent a new universality class. Some of its
characteristics are reminiscent of directed invasion percolation but
the behavior is qualitatively different.  An important feature of
directed invasion percolation%
\cite{obukhov:dirperc} %
is that a local advance of ``fluid'' is not retarded by any elastic
interaction with the rest of the fluid (unlike in models with
elasticity representing surface tension). Our model goes further
because the advancement of a local structure ahead of the rest of the
particles is actively favored: as the bottom of a drainage tree moves
down it collects more particles from above and hence is more likely to
advance further under the ``weight'' of particles from above.

The exponents $\nu$, $\alpha$ and $\Gamma$ have direct analogs in
directed percolation, where they take the values%
\cite{stauffer:introperc} %
\begin{eqnarray}
  \nu_{\text{dir-perc}}&\approx&1.73 \\
  \alpha_{\text{dir-perc}}&\approx&0.63 \\
  \beta_{\text{dir-perc}}&\approx&0.28 
  \label{eq:dirpercexp}
\end{eqnarray}
with $\beta_{\text{dir-perc}}$ analogous to our $\Gamma$.  These are
significantly different from our values. Indeed, even in mean field
theory, which has only been analyzable for adiabatic forcing below
threshold, our system is already different from directed percolation.

An important question in investigating a new class of critical
phenomena is how many independent exponents there are. We will argue
that the scaling exponents of the current density and of the drainage
basin should be related,
\begin{equation}
  \beta=1+\Gamma.
\label{eq:scalesudden}
\end{equation}
If in fact this is an equality, as is supported by the numerics and
analytic arguments, then there would be only three independent
exponents ($\beta$, $\nu$ and~$\alpha$) as for directed
percolation. In fact, we will argue that the exponent~$\alpha$ should
be exactly given by
\begin{equation}
  \alpha=\frac{1}{2},
\end{equation}
(consistent with the numerics) which would imply only {\em two
  independent exponents\/} in our system. These issues are closely
tied to the relationship between the behavior with adiabatic versus
sudden forcing.

With adiabatic forcing only one particle moves at a time when the
system is below threshold. This means that there are no particle
collisions---i.e., two particles {\em never\/} arrive at a site at the
same time---and all particles move through primary outlets, i.e.,
there are no splits below threshold. In contrast, for sudden forcing
many particles are released at once and then move down the lattice
until they find traps. Even below threshold (where there is no
steady-state current) there will be a transient flow with many
collisions between particles.  These collisions cause particles to use
secondary outlets and explore sites which would otherwise not have
been reached. This means extra traps are found and the threshold force
is increased.  Indeed we find (Sec.\ref{sec:fsspc}) critical forces
for the two cases that
%parameterized by $p$ (Sec.~\ref{sec:forceic})
%\begin{equation}
%  p_{\text{c,a}}=0.299\pm0.0005  
%\end{equation}
%for adiabatic forcing and
%\begin{equation}
%   p_c=0.3133\pm0.0003
%\end{equation}
%for sudden forcing. The values 
are quite close but clearly different.  But a more subtle question is
whether the collisions change the universality class. 

If the only effect of the collisions is for a finite fraction of the
transient moving particles to move along different outlets then the
type of forcing should not effect the universality class, as argued in
Sec.~\ref{sec:adiab}.  For adiabatic forcing the scaling relation
\begin{equation}
\beta_a=1+\Gamma_a,
\end{equation}
obtained by considering the effects of extra added particles,
should be exact (Sec.~\ref{sec:adiab}). If the effect of the
collisions is unimportant asymptotically then the two types of forcing
are in the {\em same universality class\/} and the scaling relation
[Eqs.~\ref{eq:scalesudden}] holds.

To test whether the universality classes are different one might hope
to measure different critical exponents for the adiabatic case.
Unfortunately direct simulations above threshold are difficult
because, in principle, the system must be allowed to reach a
steady-state before adding each new particle when above threshold.
Thus we are restricted to scaling below threshold. From this we find
\begin{equation}
\nu_{a}=1.60\pm0.05  
\label{eq:nua}
\end{equation} 
which overlaps with the value for the sudden case. We also find
$\Gamma_{a}/\nu_{a}=0.29\pm0.02$ which when combined with the above
value for $\nu_{a}$ leads to
\begin{equation}
  \Gamma_{a}=0.49\pm0.04,
\label{eq:gammaa}
\end{equation}
which again is well within the error bars of the sudden forcing value.

On the basis of the data and the physical arguments of
Sec.~\ref{sec:adiab} we conjecture that
\begin{equation}
  \beta=1+\Gamma=1.51\pm0.03
\end{equation}
is exact for {\em both sudden and adiabatic forcing\/} and that they are in
the {\em same universality class}.

This conclusion, and the presence of two diverging correlation
lengths in the flowing phase with only the longer one manifesting the
effects of collisions suggests that at the critical points, collisions
between moving particles and the subsequent divergence of their paths
are {\em dangerously irrelevant}. The collisions are clearly crucial
at long times above threshold: if they did not occur at all the moving
particles would eventually collect together on a single drainage tree
with a diverging local current density. But near threshold the
physics of the collisions appears to manifest itself only on very long
length (or time) scales, $\xi_n$. Most of the physical properties near
threshold---the mean current density, the drainage basin density, the
statistics of the drainage trees etc---do {\em not\/} depend on the
collisions. In Sec.~\ref{sec:intermed} we will use the behavior in the
collisionless regime above threshold on lengths scales,~$L$, in the
range
\begin{equation}
  \xi_f\ll L \ll \xi_n,
\end{equation}
to argue that the exponent~$\alpha$ should be exactly~$1/2$. This will
also make natural contact with the continuous fluid river model system
studied earlier%
.\cite{narayan:nlflow}

%%%%%%%%%%%%%%%%%%%%%%%%%%%%%%%%%%%%%%%%%%%%%%%%%%%%%%%%%%%%%%%%%%%
\section{Model}
\label{sec:modelimp}

We now define the details of the model studied here.  The model
consists of particles moving on a lattice. The (quenched) randomness
of the medium is represented by how many particle can be held
(trapped) at a site before further particles ``overflow'' to the next
site, and a local rule for distributing overflowing particles among
nearest neighbor downhill sites.  Our main focus is on two dimensional
systems and we consider motion on a square lattice oriented as shown
in Fig.~\ref{fig:paths2}. The applied force acts along one of the
diagonals and particles can move to nearest neighbor sites along the
two possible downhill directions, so that each lattice site has two
inlets and two outlets.

The way in which particles move out of a site where the local capacity
is exceeded depends on the exact variant of the model under
consideration. In this paper we use only the simplest case (the
``one-deep model'') which is described in detail in the next section.
However, conceptually it is useful to think more generally in terms of
choosing a {\em barrier height\/} for each of the two outlets from
some distribution. When the number of particles is less than the lower
outlet barrier, then no particles leave.  If the number of particles
at the site exceeds only one of the outlet heights then all of the
particles above the lower outlet's height leave by that outlet (which
we call the ``primary outlet'').  If the number of particles exceeds the
heights of both outlets then the number above the higher outlet is
divided in some way between the two outlets. 

It is convenient to work with integer outlet barriers that are the
least integer greater than the continuous valued barrier height.
Increasing the driving force,~$F$, corresponds to lowering all of
barrier heights uniformly, resulting in some fraction of the integer
barrier heights decreasing by one. This is equivalent to adding a
particle to some fraction of the sites. However, the particle sites
are not chosen completely independently, since decreasing all barriers
together means that a new particle will not appear a second time on
any site until a new one has appeared on every site.  For most of our
purposes this difference will be unimportant.

%%%%%%%%%%%%%%%%%%%%%%%%%%%
\subsection{One deep model}
\label{sec:onedeep}

\begin{figure}
  \begin{center}
    \leavevmode
    \epsfxsize=4truein
    \epsfbox{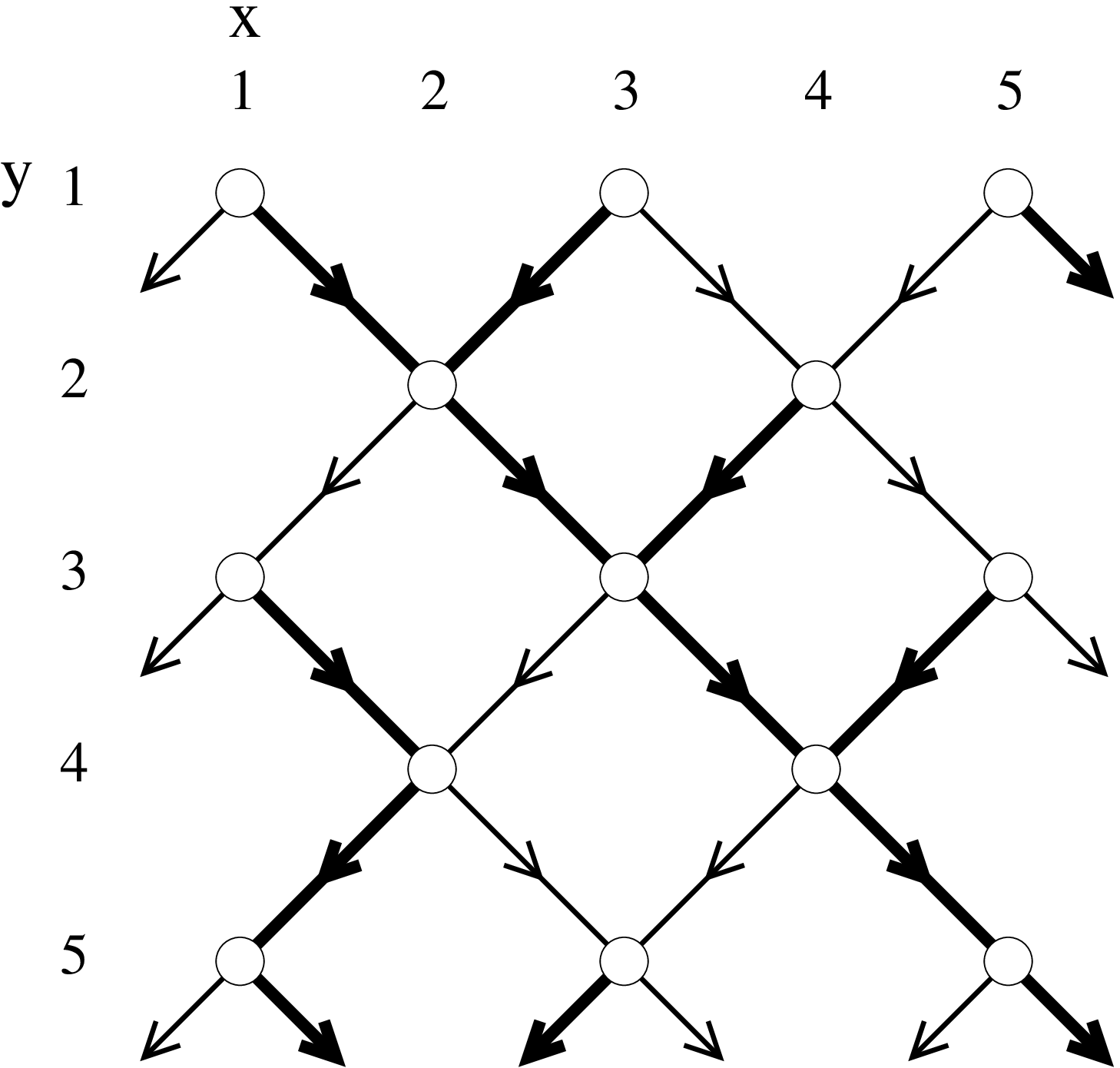}
  \end{center}
  \caption{Lattice sites with bold lines
  indicating primary outlets. Forcing is applied in the direction of
  increasing~$y$}
  \label{fig:paths2}
\end{figure}

The simplest version of this model was introduced in
Ref.~\onlinecite{watson:disc}, it allows at most one particle to past
through each outlet in one time step (hence ``one-deep'').  The
horizontal rows of the rotated square lattice of Fig.~\ref{fig:paths2}
are denoted by~$y=1,2,3\ldots$, numbering from the top down, and the
sites on row~$y$ by~$(x,y)$, with~$x+y$ an even integer. The (integer)
number of particles~$n(x,y,t)$ on site~$(x,y)$ at time~$t$ is measured
relative its capacity, i.e., to the lower integer outlet barrier so
that~$n<0$ implies that the site is unsaturated, i.e., another
particle can be trapped while $n=0$ implies the site is saturated and
$n>0$ that the site will overflow.

If~$n(x,y,t)=1$ then at time~$t+1$, one particle is moved to the site
in the next lower row to which the randomly chosen (but fixed) primary
outlet of the site connects, i.e., to one of the sites~$(x\pm1,y+1)$.
If~$n(x,y,t)=2$ then at time~$t+1$, one particle is moved to {\em
  each\/} of the two sites~$(x\pm1,y+1)$.  These rules ensure that if
we start with all $n(x,y,t=0)\leqslant2$ and update all sites
simultaneously then no more than one particle passes through any
outlet at any time step. The model is further simplified if we also
restrict unsaturated sites (traps) to a capacity of one added particle
so that
\begin{equation}
 -1\leqslant n(x,y,t)\leqslant 2.
\end{equation}

A variety of lattice sizes were chosen for numerical simulations.
Lattice lengths ($L$) are measured on the $y$-scale of
Fig.~\ref{fig:paths2}, widths ($W$) on the $x$-scale. A directed
random walker on the lattice moving along outlets has a diffusion
constant of $1/2$, i.e., the mean square displacement of a walk of
length $y$ grows as $\left<\Delta x(y)^2\right>\approx y$ for
large~$y$. As discussed in Sec.~\ref{sec:widthscaling}, lattice widths
were generally chosen scaled with the square-root of the length. Wide
systems of $W=16L^{1/2}$ were used for most measurements so that $W\gg
L^\alpha$ for any reasonable value of $\alpha$. This works well for
measurements of the current and basin fraction which become
independent of system width for large widths, but is not suitable for
measurements of $\Pi$ which approaches unity for any value of $F$ as
the system is made very wide.

All of the simulations use toroidal periodic boundary conditions. Any
particle leaving the left edge of the system appears at the right hand
side and any particle leaving the bottom row is replaced in the
corresponding site in the top row.

%%%%%%%%%%%%%%%%%%%%%%%
\subsection{Forcing and initial conditions}
\label{sec:forceic}

In this paper our primary focus is ``sudden'' forcing. The lattice is
taken to begin with all sites below capacity and then the force is
suddenly increased to the value of interest. This increase causes
sites to overflow releasing particles at many sites simultaneously.
The external force is then held constant and the dynamics of particle
flow allowed to proceed.

This sudden forcing is realized by simply using an ($F$~dependent)
initial condition with particles placed randomly at each site. Only
the number of particles at each site relative to its capacity ($n$) can
play a role. We choose this independently for each site, from a
distribution that, in general, will have weight for both positive
(excess) and negative (unsaturated) values. Our standard choice is for
the initial number of particles at each site [$n(x,y,0)$] to be either
$+1$ or $-1$ with independent probability~$p$ for each site to be
above threshold, i.e.,
\begin{equation}
  n(x,y,0)=\left\{
\begin{array}{cl}
  1
  &\quad\text{with probability~$p$}\\
  -1
  &\quad\text{with probability~$1-p$}
\end{array}\right.
\end{equation}
The parameter $p$ for this particular initial condition is a linear
function of the force $F$ used in more general discussions.

If we are studying the steady-state behavior above threshold, then it
is sometimes convenient to only allow sites to start at capacity or
above. This removes the site-filling transients and simplifies the
behavior; it will be used to study the histograms that show
convergence to the channel network.

Although sudden forcing is used for most of the simulations, we will
also contrast the behavior with that of systems with adiabatic
forcing.  Adiabatic increase of the force means that the force is
increased very slowly up to a final value $F$. Ideally the increase is
so slow that only one site overflows at a time and any released
particle reaches an unsaturated site and is trapped before another
particle is released.  Obviously this is problematic above threshold
when particles continue to flow indefinitely and interactions between
particles must be included in some manner. Our simulations with
adiabatic forcing will not extend above threshold. Below threshold,
the absence of collisions between particles and the consequent flow of
all of them thought primary outlets, enables an efficient algorithm to
be used that is described in Ref.~\onlinecite{narayan:nlflow}.

%%%%%%%%%%%%%%%%%%%%%%%%%%%%%%%%%%%%%%%%%%%%%%%%%%%%%%%%%%%%%%%%%%%
\section{Dynamic Critical Behavior}
\label{sec:crit}

In this section, we present and analyze numerical simulations on the
one-deep model near to the sudden threshold, focusing on the critical
behavior. We make extensive use of finite size scaling analysis.

%%%%%%%%%%%%%%%%%%%%%%
\subsection{Finite size scaling}
\label{sec:scalingfuncs}

In conventional isotropic systems it is often useful to use an
appropriate dimensionless quantity that is expected to approach a
non-trivial constant value at criticality. For percolation, the
crossing probability of finite size blocks is a convenient choice. A
roughly analogous quantity in our case is the fraction of systems with
a steady-state current~$\Pi(L,W,p)$.  In a large anisotropic system of
length~$L$ and width~$W$ anisotropic finite size scaling%
\cite{binder:anisofss} %
suggests that~$\Pi$ is a function only of the ratios of the
correlation lengths to the system dimensions
\begin{equation}
  \Pi(L,W,p)\approx\Pi(L/\xi_{f},W/\xi_{f_\perp}),
\end{equation}
near the critical point.

If the correlation length in the direction perpendicular to the flow
is given by $\xi_{f\perp}\propto\xi_f^{\alpha}$ then we can use an
equivalent form
\begin{equation}
  \Pi(L/\xi_f,W/L^{\alpha}).
  \label{eq:finform}
\end{equation}
For anisotropic systems it is thus important to know the correct value
of $\alpha$ when choosing the sizes of the systems if we are to use
how $\Pi$ changes with $L$ to obtain useful information.

Ideally one could perform simulations at a fixed value
of 
\begin{equation}
  w\equiv \frac{W}{L^{\alpha}}.
\end{equation}
Matching finite size scaling expectations to the critical forms of $J$
and $B$ described in Sec.~\ref{sec:critintro} for an infinite
system, we expect the scaling forms for a system of length $L$ and
fixed~$w$ to be
\begin{eqnarray}
  \Pi(L,W,p)   &\approx&                 \hat{\Pi}_w   (L/\xi_f),\\
  J(L,W,p)     &\approx&L^{-\beta/\nu} \:\hat{J}_w     (L/\xi_f),\\
  B(L,W,p)&\approx&L^{-\Gamma/\nu}\:\hat{B}_w(L/\xi_f).
\end{eqnarray}
The $w\rightarrow\infty$ limits of the current and basin fraction
scaling functions should be $w$-independent functions that provide a
useful way of making $\alpha$-independent measurements with wide
systems.  However $\hat{\Pi}_w(L/\xi_f)$ approaches~1 as
$w\rightarrow\infty$ for any value of $L/\xi_f$ which is not useful.

With the expectation that
\begin{equation}
\xi_f\propto \frac{1}{(p-p_c)^{\nu}}  
\end{equation}
we can write these
as
\begin{eqnarray}
  \Pi(L,W,p)       &\approx&
  \tilde{\Pi}_w\left((p-p_c)L^{1/\nu}\right) 
  \label{eq:pisf}\\
  J(L,W,p)         &\approx&
  L^{-\beta/\nu} \:\tilde{J  }_w\left((p-p_c)L^{1/\nu}\right)
  \label{eq:jsf}\\
  B(L,W,p)    &\approx&
  L^{-\Gamma/\nu}\:\tilde{P  }_w\left((p-p_c)L^{1/\nu}\right)
  \label{eq:bsf}
\end{eqnarray}
where $\tilde{\Pi}_w$, $\tilde{J}_w$ and~$\tilde{P}_w$ are the scaling
functions for the particular value of~$w$. As will be discussed later,
the consideration of the long distance behavior above threshold
suggests that $\alpha$ should be equal to~$1/2$---essentially from
random walk scaling of the primary outlet paths.  We thus first, in
the next three subsections, use the value $\alpha=1/2$ and the above
scaling forms to infer the other exponents. Although the fits are good
they may be biased by the choice of~$\alpha$.

However, an alternative set of single variable scaling forms can be
used {\em without\/} knowing $\alpha$, by working at $p=p_c$. Putting
$\xi_f=\infty$ in the form of~Eq.~(\ref{eq:finform}) yields the scaling
forms
\begin{eqnarray}
  \Pi(W,L,p_c)       &\approx&
  \tilde{\Pi}_c(W/L^{\alpha})
  \label{eq:piwsf}\\
  J(W,L,p_c)         &\approx&
  L^{-\beta/\nu} \:\tilde{J  }_c(W/L^{\alpha})
  \label{eq:jwsf}\\
  B(W,L,p_c)    &\approx&
  L^{-\Gamma/\nu}\:\tilde{P  }_c(W/L^{\alpha}).
  \label{eq:bwsf}
\end{eqnarray}
where $\tilde{\Pi}_c$, $\tilde{J}_c$ and~$\tilde{P}_c$ are single
variable scaling functions. These forms will be used, together with a
measurement of $p_c$, to determine $\beta/\nu$ and $\Gamma/\nu$ and to
try and measure $\alpha$ in Sec.~\ref{sec:widthscaling}.

%%%%%%%%%%%%%%%%%%%%%%%
\subsection{Scaling of~$\Pi$}
\label{sec:fsspc}

\begin{figure}
  \begin{center}
    \leavevmode
    \epsfxsize=4truein
    \epsfbox{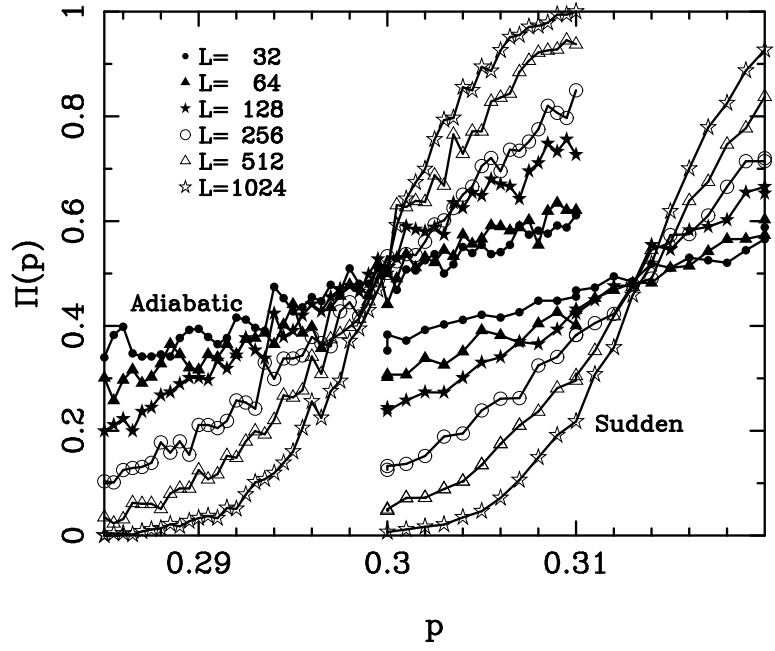}
  \end{center}
  \caption{Fraction of systems that have
  a steady-state current for sudden and adiabatic increase of the
  force. Data use the same set of system sizes with widths
  $W=4L^{1/2}$ and using 512 samples for each.}
  \label{fig:bothcross}
\end{figure}

The finite size scaling form of~$\Pi$ [Eq.~(~\ref{eq:pisf})] allows us
to measure~$p_c$ and the exponent~$\nu$ (at least if we know~$\alpha$)
We first identify $p_c$ by noting that $\Pi(L,p=p_c)=\tilde{\Pi}_w(0)$
which should be a constant independent of $L$. This means that curves
of $\Pi(p)$ for different systems sizes $L$ should all intersect at
$p=p_c$.  Fig.~\ref{fig:bothcross} shows measurements of $\Pi(p,L)$
for both adiabatic and sudden dynamics. Each set of curves intersects
quite accurately at a single point.

% cat j5-picross.plt j5.dat k5-picross.plt k5c.out k5d.out k5e.out

\begin{figure}
  \begin{center}
    \leavevmode
    \epsfxsize=4truein
    \epsfbox{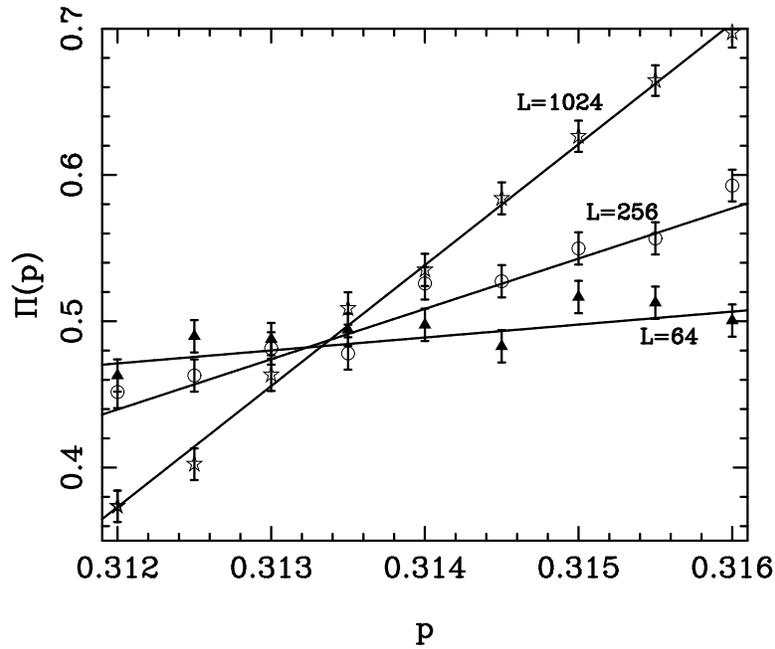}
  \end{center}
  \caption{Close up of $\Pi(p)$ curves
  for selected lengths with sudden forcing. Error bars are from
  measured sample variance. Crossing point is estimated as
  $p_c=0.3133\pm0.0003$.}
  \label{fig:picrosszoom}
\end{figure}

The data for $\Pi(p)$ was taken using a series of systems of width
$W=4L^{1/2}$. The prefactor of~4 was chosen to place the crossing
point near $\Pi=1/2$ where repeated simulations converge most quickly
to an estimate of $\Pi$ and where the crossing is most easily seen. By
looking closer at the crossing of curves for different size systems
(as in Fig.~\ref{fig:picrosszoom}) we determine $p_c=0.3133\pm0.0003$
for sudden forcing and $p_{c,a}=0.2990\pm0.0005$ for adiabatic forcing.

% jplot j3-picross.plt j3.dat

\begin{figure}
  \begin{center}
    \leavevmode
    \epsfxsize=4truein
    \epsfbox{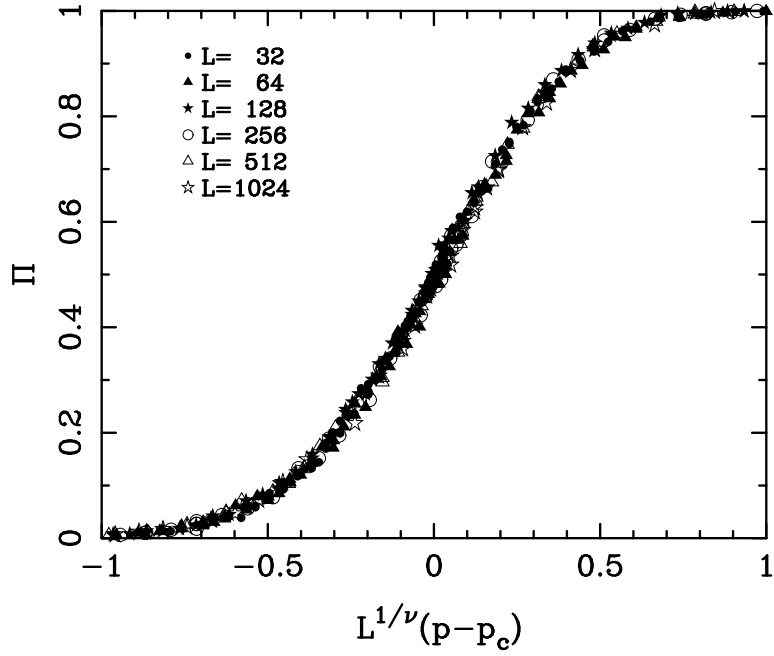}
  \end{center}
  \caption{$\Pi$ finite size scaling with
  $p_c=0.3133$ and $\nu=1.62$ for systems of width $W=4L^{1/2}$}
  \label{fig:pifss}
\end{figure}

With this value of $p_c$ we can now examine how the form of $\Pi(p)$
changes with the system size. The finite size scaling hypothesis
suggests that a plot of $\Pi$ against $L^{1/\nu}(p-p_c)$ for different
lengths should collapse onto a single curve.  Such a plot is shown in
Fig.~\ref{fig:pifss}. By varying $\nu$ and comparing how well the data
collapses we estimate $\nu=1.62\pm0.04$.

% jplot ja-pifss.plt ja-new.dat (less 16)

\begin{figure}
  \begin{center}
    \leavevmode
    \epsfxsize=4truein
    \epsfbox{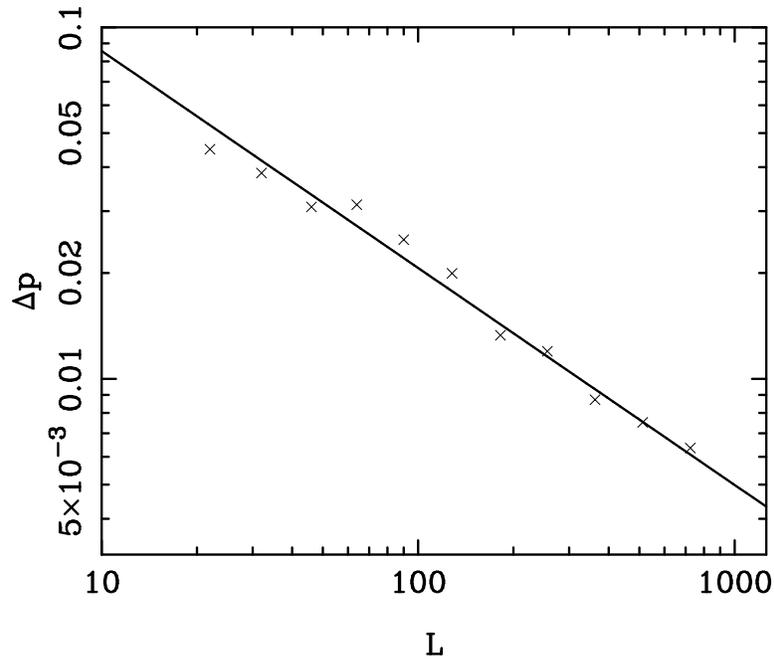}
  \end{center}
  \caption{Width ($\Delta p$) of peak in
  $\partial\Pi/\partial p$ for systems of different lengths, and
  widths~$W=4L^{1/2}$. Line has slope $-1/\nu$ with
  $\nu=1.62$.}
  \label{fig:deltap}
\end{figure}

Another estimate of $\nu$ comes from the width of the region over
which $\Pi(p)$ is changing%
.\cite{stauffer:introperc} %
If we differentiate $\Pi(p)$ with respect to $p$ we get a curve which
peaks near $p=p_c$ and has width proportional to $L^{-1/\nu}$. The
gradient of a log-log plot of the width of the peak ($\Delta p$)
against $L$ should be $-1/\nu$. The width can be obtained from simple
integrals of the $\Pi(p)$ data,
\begin{equation}
  (\Delta p)^2\equiv
  \int_0^1\left(\frac{d\Pi}{dp}\right)(p-\bar{p})^2\:dp 
%         &=1-\bar{p}^2-2\int_0^1 p\Pi(p)\:dp
\label{eq:parts1}
\end{equation}
where
\begin{equation}
  \bar{p}\equiv\int_0^1 \left(\frac{d\Pi}{dp}\right) p \: dp
%  =1-\int_0^1 \Pi(p) \: dp.
\label{eq:parts2}
\end{equation}
is the center of the peak. No differentiation is necessary if
integration by parts is used in Eqs.~(\ref{eq:parts1})
and~(\ref{eq:parts2}), which reduces the noise in the calculations.
Fig.~\ref{fig:deltap} shows $\Delta p$ together with a line of the
expected slope using the value of $\nu=1.62$ from above.  Although the
data points have the expected linear form they do not give an accurate
value for the slope. Linear regression gives a rather imprecise
estimate, $\nu=1.69\pm0.08$.

%% Data is from j5 and j6 using two different fitting methods.

Our best estimate of $\nu$ is thus $\nu=1.62\pm0.04$. This will turn
out to be the least accurately determined of all our exponents.  This
is primarily because measuring $\Pi$ is much less efficient than
measuring $B$ or $J$. A single run for a large system gives a
good estimate of $B$ and $J$, but only provides a zero or one
value for $\Pi$.  Many systems must be averaged over to get a good
estimate of the fraction that contain moving particles. A further
advantage of measuring $B$ and $J$ is that wide systems can be
used which effects more averaging from the larger system and gives
results that are less sensitive to the value of~$\alpha$. We thus turn
to these quantities.

%%%%%%%%%%%%%%%%%%%%%%%
\subsection{Current density}
\label{sec:fssj}

\begin{figure}
  \begin{center}
    \leavevmode
    \epsfxsize=4truein
    \epsfbox{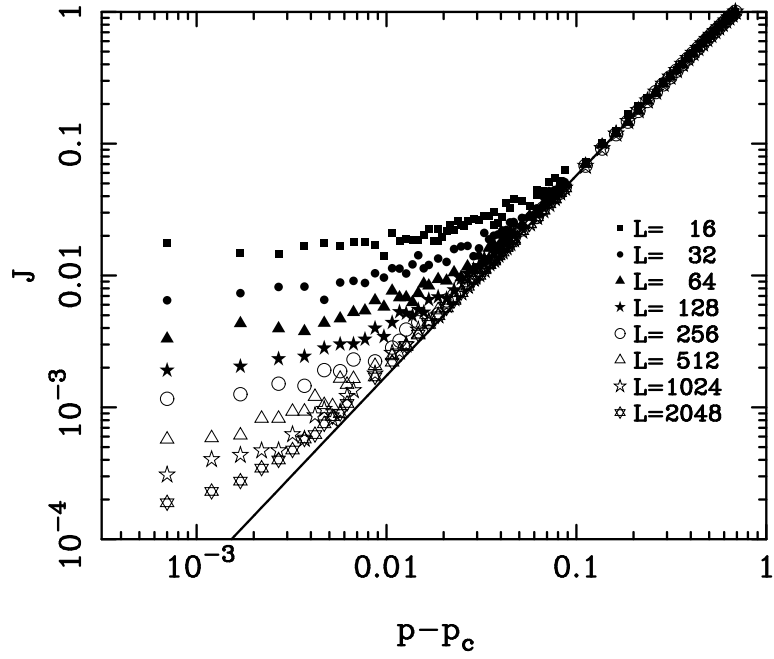}
  \end{center}
  \caption{Log-log plot of $J$ versus $(p-p_c)$ for
  $p>p_c$ with
  $p_c=0.3133$. Line shown has slope~$\beta=1.53$}
  \label{fig:logj}
\end{figure}

In an infinite system we expect $J\sim (p-p_c)^\beta$. In a finite
size system this should only hold when $\xi_f(p)\ll L$. Data from
simulations of different lengths (and large widths $W=16L^{1/2}$) are
shown in Fig.~\ref{fig:logj} which exhibits the form expected---the
data from each system size follows a straight line until $\xi_f(p)$ is
of order~$L$. Estimating the slope for an infinite system from this
figure gives $\beta=1.53\pm0.03$.

% was cat i1-ivlog.plt i1b-i2-i3.dat curve-beta.plt
% jplot ia-logj.plt ia.dat ia-logj-curve.plt

\begin{figure}
  \begin{center}
    \leavevmode
    \epsfxsize=4truein
    \epsfbox{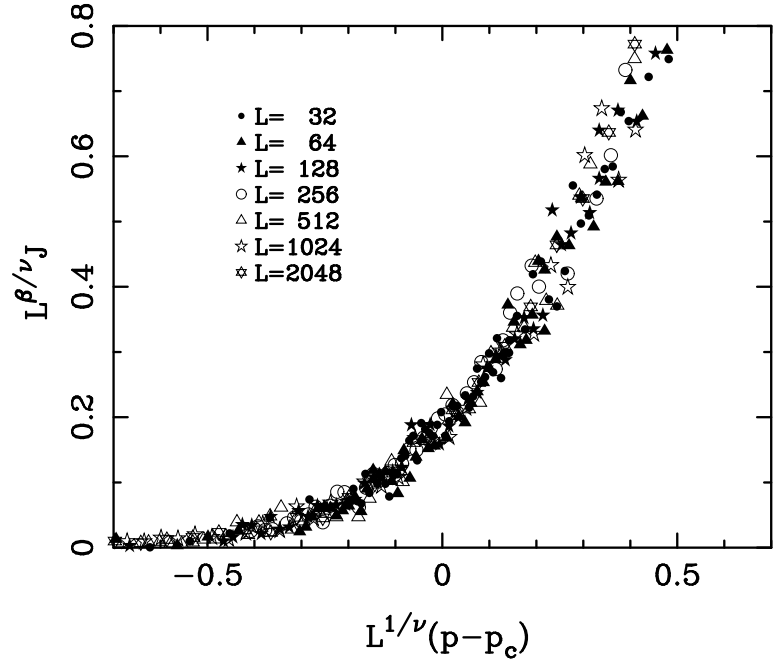}
  \end{center}
  \caption{Finite size scaling plot
  for $J$ using $p_c=0.3133$, $\nu=1.62$ and $\beta=1.53$.}
  \label{fig:jfss}
\end{figure}

We can also test the finite size scaling form for the current,
Eq.~(\ref{eq:jsf}), as shown in Fig.~\ref{fig:jfss}.  Varying the
value of $\beta$ suggests that $\beta/\nu=0.90\pm0.03$.

% was cat i1-ivfss.plt i1b-i2-i3-vbig.dat
% jplot ia-jfss.plt ia.dat

\begin{figure}
  \begin{center}
    \leavevmode
    \epsfxsize=4truein
    \epsfbox{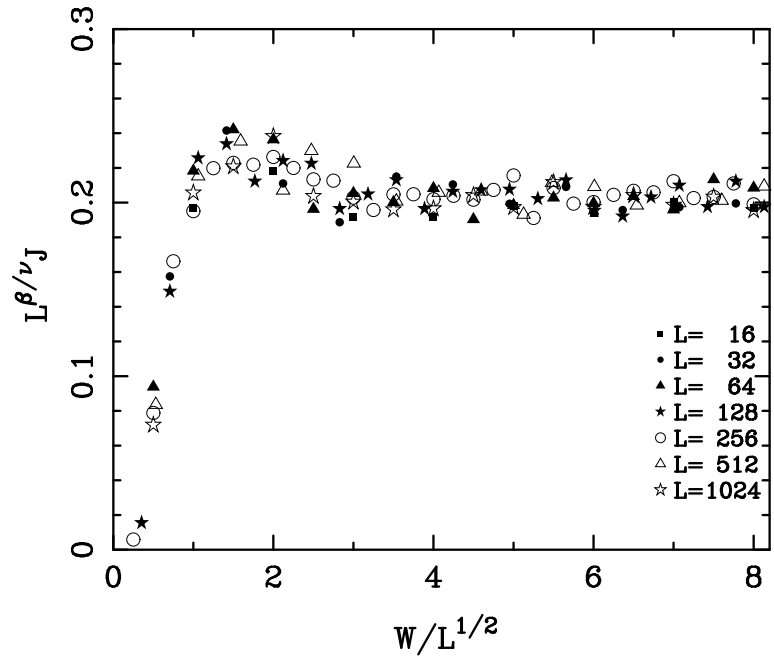}
  \end{center}
  \caption{Scaling of current at $p=0.3133$ with
  sample size, using $\beta/\nu=0.96$}
  \label{fig:widthj}
\end{figure}

An alternative way to determine $\beta/\nu$ is to use the width
scaling form of $J$ [Eq.~(\ref{eq:jwsf})].  This gives a value for
$\beta/\nu$ without requiring $\nu$ to be known ($p_c$ is needed but
is known more accurately).  Because the current becomes
approximately independent of the width once $W\gtrsim 2L^{1/2}$ (see
Fig.~\ref{fig:widthj}) this gives a rather sensitive test.  Working
at $p_c=0.3133$ this gives $\beta/\nu=0.96\pm0.01$. With
$\nu=1.62\pm0.04$ this translates to $\beta=1.55\pm0.04$.

%The scaling function $\tilde{\Pi}_c(x)$ has a small peak at
%$x\approx1.5$, one might hypothesize that this corresponds to the
%natural size of a drainage structure of length $L$.

% cat j4-iv.plt j4.dat >widthj.plt

Combining this~$\beta/\nu=0.96\pm0.01$~value from width scaling with
our best value $\beta=1.53\pm0.03$ gives $\nu=1.60\pm0.04$ which
agrees well with the value of~$1.62\pm0.04$ determined independently
in Sec.~\ref{sec:fsspc}.

%%%%%%%%%%%%%%%%%%%%%%%
\subsection{Drainage basin}
\label{sec:fssbs}

\begin{figure}
  \begin{center}
    \leavevmode
    \epsfxsize=4truein
    \epsfbox{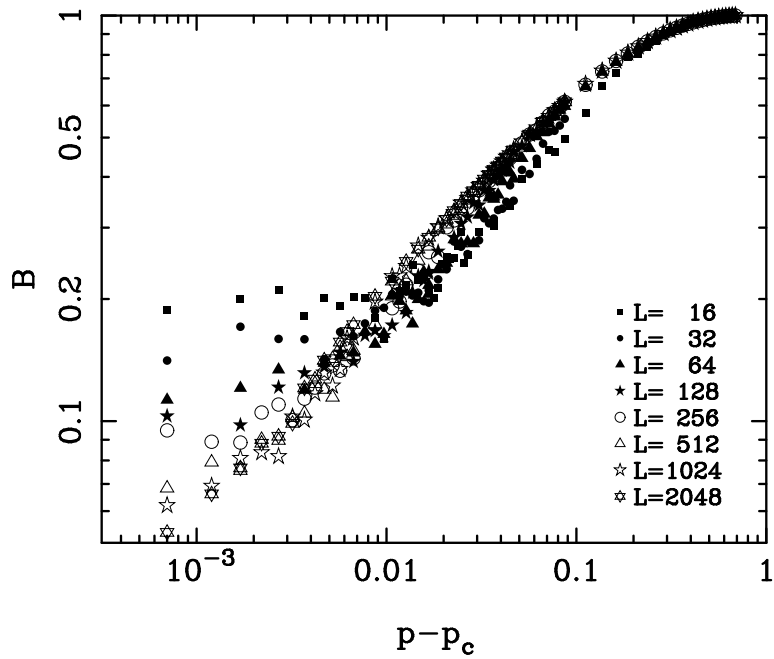}
  \end{center}
  \caption{Log-log plot for basin
  density.}
  \label{fig:logb}
\end{figure}

A direct plot of the basin fraction against $p-p_c$
(Fig.~\ref{fig:logb}) is not as well-behaved as the equivalent plot
for the current. The range of ordinate values is much smaller and the
deviation away from the power law is of opposite sign at $p-p_c$ small
and $p-p_c$ large. These factors make determining the asymptotic slope
rather difficult.  The largest sizes are approximately linear over one
and a half decades of $p-p_c$ (compared with almost two and a half
decades for~$J$)%
.\cite{footnote:unclear} %

% jplot ia-logb.plt ia.dat #ia-logb-curve.plt 

\begin{figure}
  \begin{center}
    \leavevmode
    \epsfxsize=4truein
    \epsfbox{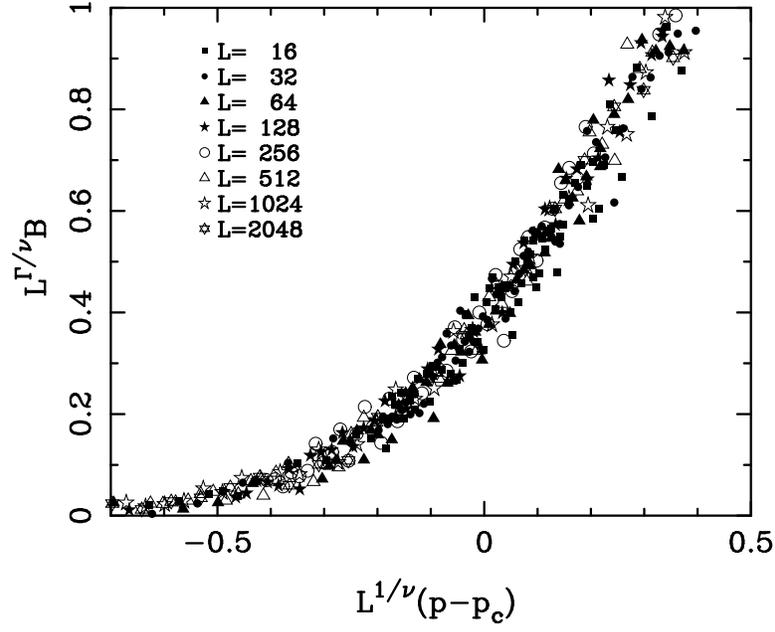}
  \end{center}
  \caption{Finite size scaling plot using
  $p_c=0.3133$, $\Gamma/\nu=0.29$ for the basin
  fraction~$B$.}
  \label{fig:pinffss}
\end{figure}

\begin{figure}
  \begin{center}
    \leavevmode
    \epsfxsize=4truein
    \epsfbox{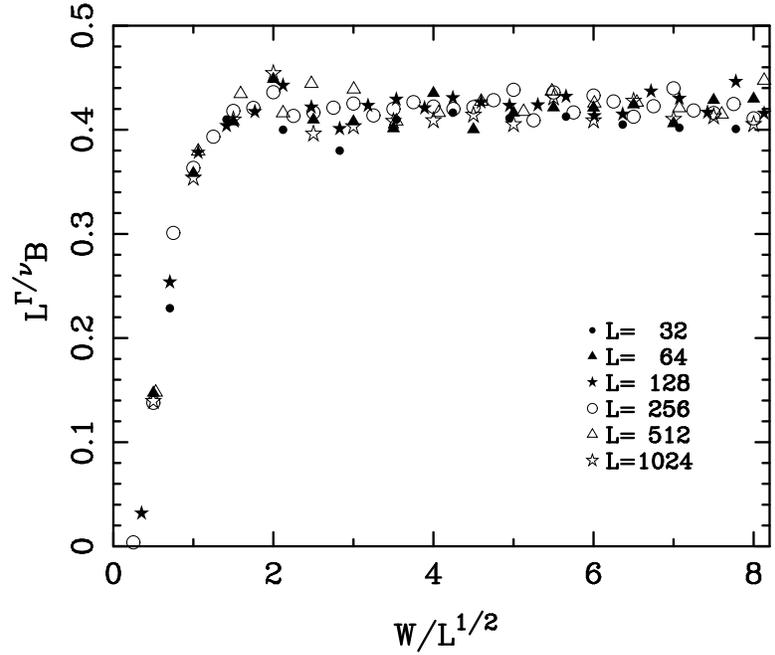}
  \end{center}
  \caption{Scaling of drainage basin fraction at
  $p=0.3133$ with sample size using $\Gamma/\nu=0.302$}
  \label{fig:widthb}
\end{figure}

Finite size scaling can also be used for the drainage basin and gives
$\Gamma/\nu=0.29\pm0.02$ (Fig.~\ref{fig:pinffss}). Varying the width
at $p_c$ gives $\Gamma/\nu=0.302\pm0.005$. With $\nu=1.62\pm0.04$ this
translates to $\Gamma=0.49\pm0.02$ (see Fig.~\ref{fig:widthb}).

% was cat i1-bsfss.plt i1b-i2-i3-big.dat >pinffss.plt
% jplot ia-pinffss.plt ia.dat

% cat j8-bs.plt j8-0.3135.dat

%%%%%%%%%%%%%%%%%%%%%%
\subsection{Anisotropic scaling}
\label{sec:widthscaling}

As explained above in Sec.~\ref{sec:scalingfuncs}, it is important to
know the value of $\alpha$ in order to choose consistent system sizes
for simulation. An analogous situation occurs in quantum Monte Carlo
situations if the dynamic scaling exponent is not known%
.\cite{young:altenbergbrp} %
In both cases
determining the anisotropy scaling exponent accurately is quite
difficult.

So far in this section we have assumed that the width of the
structures in the system scales as the square-root of their length,
i.e., that $W\propto L^{\alpha}$ with $\alpha=1/2$---although this
value is not as crucial for the wide, $W=16L^{1/2}$, systems used for
measurements of $J$ and $B$. In this section we provide numerical
support for a value close to this.

%%For this system we believe $\alpha=1/2$ may be correct. Above
%%threshold at long scales~($\gg\xi_{n}\gg\xi_f$) the current network
%%has a characteristic transverse difference between channel splits and
%%joins of~$\sim\xi_{n}^{1/2}$. This suggests the hypothesis that
%%$\alpha=1/2$ generally. We will try and produce numerical evidence for
%%$\alpha=1/2$.  
Obtaining an accurate estimate of~$\alpha$ is rather difficult, a more
modest goal will be try to rule out a value as different from~$1/2$ as
the value for directed percolation
($\alpha\approx0.63$)%
.\cite{kinzel:dirpercfss} %

In Sec.~\ref{sec:scalingfuncs} we discussed scaling forms for
simulations at $p=p_c$ which give information on the value
of~$\alpha$; we should be able to choose the value of $\alpha$ which
provides the best data collapse with
Eqs.~(\ref{eq:piwsf})--(\ref{eq:bwsf}).  The problem with this method
is that in order to locate $p_c$ we have to perform a series of
simulations using sizes scaling with some value of $\alpha$. Choosing
the wrong $\alpha$ will give an apparent value of $p_c$ that differs
from the value obtained using the correct $\alpha$. This problem is
made more acute by the fact that $p_c$ is best determined from
measurements of $\Pi$ on narrow systems which are most sensitive to
the value of $\alpha$.

To get around this problem we performed additional sets of simulations
with different system sizes chosen with the values $\alpha=0.4$ and
$0.6$ (the prefactors $A$ in $W\approx AL^\alpha$ were chosen to give
crossings near $\Pi=0.5$, but the value of $p_c$ is insensitive to the
prefactor used).  As discussed in Sec.~\ref{sec:fsspc} we used the
crossing of the $\tilde{\Pi}_w$ function to give an estimate for the
apparent critical point for each value of $\alpha$,
$\hat{p}_c(\alpha)$, obtaining
\begin{eqnarray}
  \hat{p}_{c}(\alpha=0.40)&= 0.3150\pm0.0005,\\
  \hat{p}_{c}(\alpha=0.60)&= 0.3110\pm0.0005.
\end{eqnarray}
In Sec.~\ref{sec:crit} we found
$\hat{p}_c(\alpha=0.5)=0.3133\pm0.0003$.

\begin{figure}
  \begin{center}
    \leavevmode
    \epsfxsize=4truein
    \epsfbox{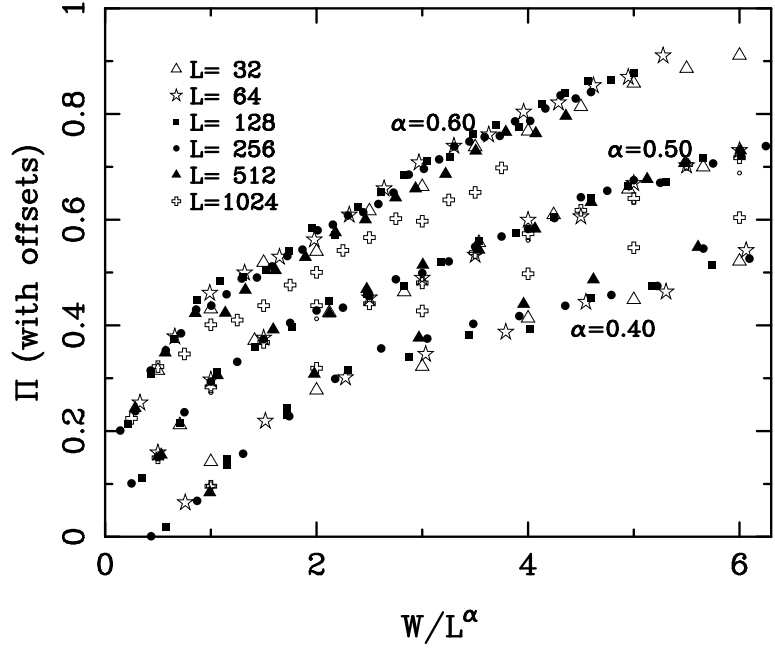}
  \end{center}
  \caption{Tests for the width scaling for
  different values of $\alpha$. Simulations run at
  $\hat{p}_c(\alpha)$. Data for $\alpha=0.6$ is offset vertically by
  $+0.2$, data for $\alpha=0.5$ is offset by $+0.1$.}
  \label{fig:alpha456}
\end{figure}

We then performed a series of simulations for different widths and
lengths with $p$~chosen to equal to the apparent critical parameter
$\hat{p}_c(\alpha)$. Plotting $\Pi$ against $W/L^\alpha$ should yield
good data collapse only for the correct value of $\alpha$
[Eq.~(\ref{eq:piwsf})]. These data are shown in
Fig.~\ref{fig:alpha456}.  The data collapse is most effective for
$\alpha=0.5$, where it appears to be limited only by the statistical
error in each point (which is approximately equal to the height of the
plotting symbols).

% cat j8-pi-0.40.plt j8-0.316.dat
% cat j8-pi-0.50.plt j8-0.3135.dat
% cat j8-pi-0.60.plt j8-0.311.dat 

\begin{figure}
  \begin{center}
    \leavevmode
    \epsfxsize=4truein
    \epsfbox{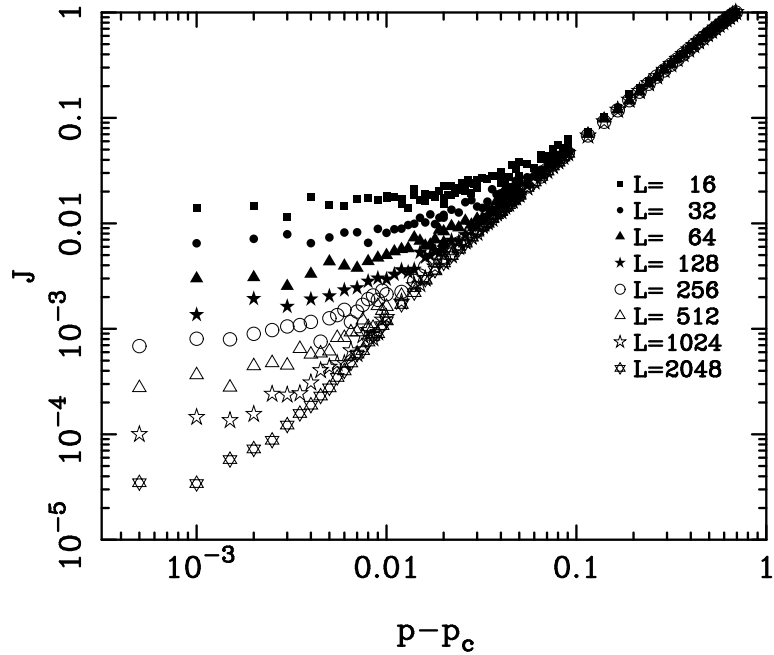}
  \end{center}
  \caption{Log-log plot of current using
  the value $p=\hat{p}_c(\alpha)$ determined from using widths scaled
  with $\alpha=0.4$. Compare with Fig.~\protect{\ref{fig:logj}}}
  \label{fig:logjwrong}
\end{figure}

It is also possible to make similar scaling plots for the current and
basin fraction. But the scaling functions for these quantities rapidly
approach a steady-state value as the width is increased and so are not
very sensitive to the value of~$\alpha$ as was seen in
Figs.~\ref{fig:widthj} and~\ref{fig:widthb}.  However, another way to
try and fix~$\alpha$ is to try to use the other values of
$\hat{p}_c(\alpha)$ to fit the data for $J$ and $B$ from simulations
on wide ($W=16L^{1/2}$) samples (which should be insensitive to the
value of $\alpha$ in a substantial range of~$\alpha$ around~$1/2$) as
$16L^{1/2}\geqslant8L^{0.6}$ even for $L$ as large as 1024). Both the
direct logarithmic plots for $J$ and $B$ and the finite size scaling
plots are significantly worsened if a value for $p_c$ as different as
$0.311$ or $0.315$ is used. For example, Fig.~\ref{fig:logjwrong}
shows how the logarithmic graph of current is changed by using the
$\alpha=0.4$ apparent~$p_c$ value of~$0.315$.

% jplot ia-logj-0315.plt ia.dat

We have provided numerical evidence to support taking $\alpha=1/2$.
Using similar criteria to our other apparent
errors, one would guess
\begin{equation}
  \alpha=0.5\pm0.05.
\end{equation}
In any case, a directed percolation-like value of~$\alpha=0.63$ seems
unlikely, Again, note that most of our measurements were done on very
wide systems so the choice of $\alpha=1/2$ should not be a significant
factor if the true $\alpha$ is slightly different.

%%%%%%%%%%%%%%%%%%%%%
\subsection{Adiabatic forcing}
\label{sec:adiab}

\begin{figure}
  \begin{center}
    \leavevmode
    \epsfxsize=4truein
    \epsfbox{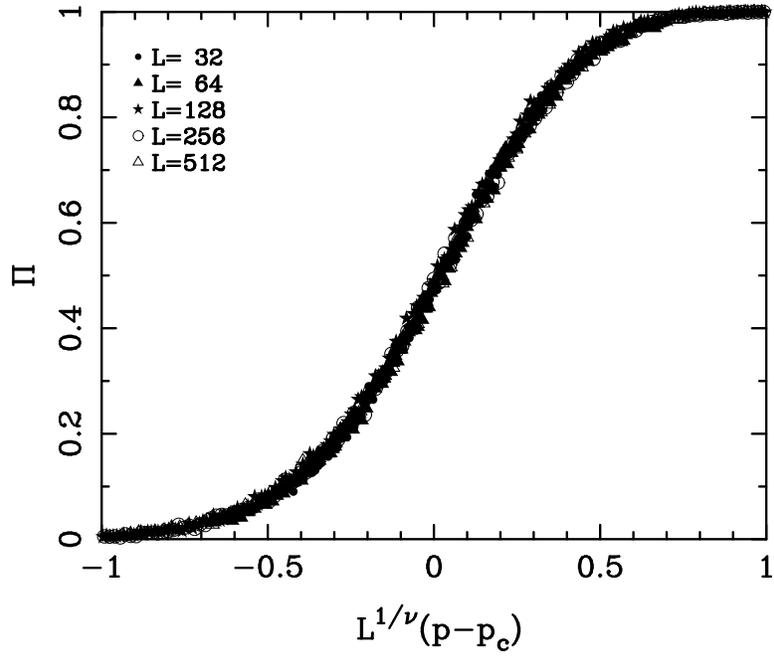}
  \end{center}
  \caption{$\Pi$ finite size scaling for
  {\em adiabatic\/} increase with $p_{c,a}=0.299$ and $\nu_{a}=1.60$
  and using narrow~($W=4L^{1/2}$) lattices}
  \label{fig:adiapifss}
\end{figure}

With adiabatic forcing, as mentioned in the Introduction, performing
good simulations above threshold is too time consuming as the system
should be allowed to reach a steady-state before each particle is
added. Thus we are limited, essentially, to studies of~$\Pi$, the
probability that some particles keep moving forever, and the drainage
basin fraction at criticality. Finite size scaling can be done for
$\Pi(p)$ for adiabatic forcing (Fig.~\ref{fig:adiapifss}). The
corresponding estimate for the exponent is~$\nu_{a}=1.60\pm0.05$. If
the scaling functions for the sudden and adiabatic cases are
superimposed from Figs.~\ref{fig:pifss} and~\ref{fig:adiapifss} they
cannot be distinguished. Using the value of the critical point,
$p_{c,a}$, determined from $\Pi$, $\Gamma_a/\nu_a$ can be obtained as
in Sec.~\ref{sec:fssbs}. These yield the exponent estimates of
Eqs.~(\ref{eq:nua}) and~(\ref{eq:gammaa}) which are the same, even
within the apparent 1-$\sigma$ errors, as those with sudden forcing.

% jplot m2-pifss.plt m2b.dat

Note that $p_c$ and $p_{c,a}$ are rather close; but given the relative
smallness of them and the observation that with sudden forcing the
number of doubly occupied sites after one time step is~$\sim p^3$, it is
not surprising that the difference is of this order. Also note that
even if the fits are done over the range
\begin{equation}
  |p-p_c|\lesssim p_c-p_{c,a}
\end{equation}
to try to separate out possible bias from the critical values being
similar, the apparent exponents do not change.

%%DSF: # of extra traps filled after 2 time steps is 
%% \approx (1/4)p^3(1-(3/2)p)\qpprox0.004
%%

%%%%%%%%%%%%%%%%%%%%%%%
\subsection{Equivalence and scaling}
\label{sec:equiv}

As mentioned in the Introduction, both our numerical results and
analytic arguments for the adiabatic forcing suggest a scaling
relation between~$\Gamma_a$ and $\beta_a$. Here we explain the
adiabatic result (following Ref.~\onlinecite{narayan:nlflow} for the
case of a continuous fluid) and give some plausibility arguments for
the sudden case which also suggest that both cases are in the same
universality class.

If we increase the force {\em adiabatically\/} by an infinitesimal
amount~$\delta F$ then a fraction $\delta F$ of all the sites will
overflow and the current will be increased by the overflow from those
sites which are on the drainage basin. The current thus increases by
$\delta J=B\:\delta F$, which, when integrated, gives the relation
\begin{equation}
  \beta_a=1+\Gamma_a.
  \label{eq:adiasc}
\end{equation}
A possible weakness in this argument is that it neglects the
anti-correlation between the extra sites that overflow and the
previous sites that have already overflowed. In the simplest version
of the model no site will overflow for a second time until all sites
have overflowed once. This effect could reduce the amount of
current induced, but should not be singular near threshold and thus
should not change the scaling law Eq.~(\ref{eq:adiasc}).

If we instead consider sudden increases of the force from an initial
value well below threshold then the argument is more complicated.  If
the force is suddenly increased to $F$, there will be a transient
current $J(F,t)$ which decays to the steady-state current
$J(F,\infty)$.  Now imagine {\em instead\/} increasing the force to a
final value of $F+\delta F$.  The initial transient current will now
be
\begin{equation}
  J(F+\delta F, t=0)=J(F,t=0)+\delta F.
\end{equation}
The extra particle density will be randomly distributed over the
lattice (except for the anti-correlation with previously overflowed
sites mentioned for the adiabatic case) and some fraction of them will
survive to join the steady-state current.  An extra particle added to
any site that is on the steady-state drainage basin will increase the
final steady-state current by one particle, as long as the extra
particle passes only through primary outlets (i.e., if it does not
collide with any other particles),  while an extra particle initially
off the drainage basin will fall into a trap and not contribute to the
steady-state current unless it has a collision with another particles
that forces one of them onto the drainage basin.  If we neglect these
collisions then we would have the same result ($\beta=1+\Gamma$) as
for the adiabatic case with the minor difference that the extra
steady-state current is not made up entirely of the new initial
particles: some new particles will fill in traps and allow particles
that would otherwise have been trapped to survive.

The complicating factor is that with sudden forcing we cannot neglect
collisions caused by the new particles.  In the adiabatic case the
only particles were the new ones with density $\delta F$ so the
collision rate was only of order $(\delta F)^2$. In the sudden case
collisions between the new particles are similarly negligible but a
collision rate of $\sim J(F,t)\delta F$ between the new particles and
the existing particles is expected. This is much larger near $F_c$ as
$J(F,t)$ does not vanish for small times. The extra collisions can
reduce the steady-state current (which will change $F_c$) but will not
induce violation of the scaling relation unless the factor by which
the steady-state current is reduced is critical. Such a singular
contribution is only likely to be produced from the long time limit of
the decaying current, but in this regime the current is small near
$F_c$ so there will be few collisions and a critical reduction seems
unlikely. This suggests that~$\beta=1+\Gamma$ holds and also that the
presence of splits does not change the universality class near
threshold. In Sec.~\ref{sec:intermed} we will be more quantitative and
explore the condition on the exponents for this argument to be valid.
We will see that it also leads to $\alpha=1/2$.

It is possible to relate the exponents we have computed to the
fractal dimension of the drainage basin on intermediate scales or of
large drainage trees below threshold. For comparing with other systems
we work in general dimensions: 1 downhill and $d-1$ horizontal.
Drainage trees of length $L$ smaller than $\xi_f$ contain a typical
number of sites $\sim L^{d_f}$.  In a system of length $L\gg\xi_f$
(with the perpendicular dimensions $\sim L^{\alpha}$) we expect the
largest drainage cluster to contain a number of sites $\sim B
L^{1+\alpha(d-1)}$. If $L\ll\xi_f$ then the largest cluster will
contain $\sim L^{d_{f}}$ sites. For these forms to match at
$L\sim\xi_f$ the hyper-scaling relation
\begin{equation}
  d_{f}=1+\alpha(d-1)-\frac{\Gamma}{\nu}
  \label{eq:hypscaling}
\end{equation}
has to hold. In mean field theory, expected%
\cite{narayan:nlflow} %
to be valid for $d>3$, it was found that $d_f=4/3$, $\nu=3/2$ and
$\Gamma=1$, although the hyperscaling relation
[Eq.~(\ref{eq:hypscaling})] will not hold except in the critical
dimension~$d=3$ for which logarithmic corrections to mean field theory
are likely. For $d=2$ and taking $\alpha=0.50\pm0.05$,
$\Gamma/\nu=0.28\pm0.02$ Eq.~(\ref{eq:hypscaling}) gives
\begin{equation}
  d_{f}=1.22\pm0.05.
\end{equation}

There are several useful bounds on the exponents. The generalized
Harris criterion for the finite size scaling correlation length for a
random system with $(d-1)$ transverse dimensions
scaling as~$L^\alpha$ is%
,\cite{harris:criterion,chayes:fsscaling} %
\begin{equation}
  \nu\geqslant \frac{2}{1+\alpha(d-1)}
\end{equation}
which, with $d=2$ and $\alpha=1/2$, means $\nu\geqslant4/3$. Our
measured value of $\nu=1.62\pm0.04$ satisfies this bound and is not
too far away from saturating it. There is also a simple bound on the
fractal dimension because the drainage trees must be at least linear,
i.e., $d_f\geqslant1$, implying
\begin{equation}
  \frac{\Gamma}{\nu}\leqslant\alpha(d-1)
\end{equation}
which is easily satisfied.

%%%%%%%%%%%%%%%%%%%%%%%%%%%%%%%%%%%%%%%%%%%%%%%%%%%%%%%%%%%%%%%%%%%
\section{Development of the Channel Network}
\label{sec:conv}

\begin{figure}
  \begin{center}
    \leavevmode
    \epsfxsize=4truein
    \epsfbox{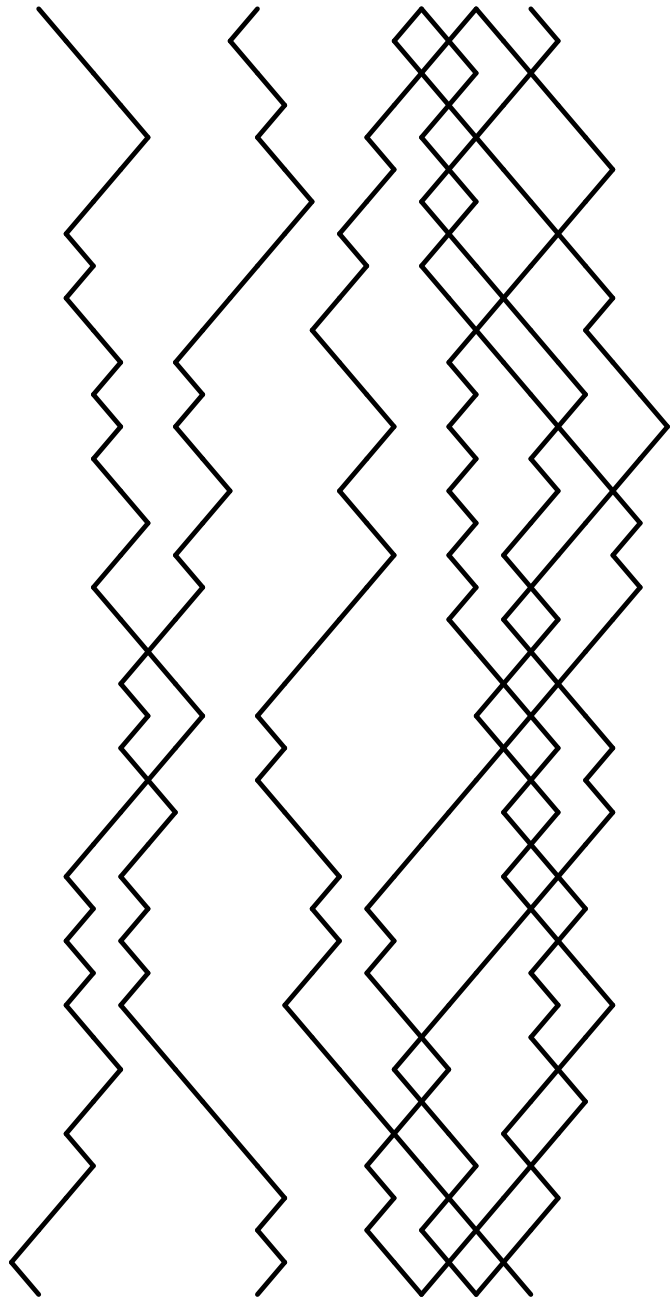}
  \end{center}
  \caption{Channel network showing which
  outlets in the lattice carry current in the steady-state}
  \label{fig:network}
\end{figure}

The channel network above threshold consists of those sites that carry
current in the steady-state (see Fig.~\ref{fig:network}).  An
important feature of our model is that in an infinite system the flow
converges (albeit slowly) to a fixed network as shown in
Ref.~\onlinecite{watson:disc}. The network is a property of the steady-state
and is asymptotically time-independent.  Which sites are on the
network depends only on the amount of steady-state current and on the
choice of primary outlets between sites. The network does not depend
on the initial placement of particles, how they enter the system or
the history of the applied force.

As discussed in Ref.~\onlinecite{watson:disc} %
these features allow the convergence to the network to be seen
directly by looking at the number of sites differing between two
copies of the same lattice.  In particular, if particles enter each
copy of the system through a different set of sites in the top row
then we predicted that the number of differing sites should decay as
$y^{-1/4}$ (for large~$y$) as the particles move down the system. We
have performed simple simulations with two such copies and measured a
value of $-0.262\pm0.005$ for the exponent which agrees well,
especially considering the difficulties in simulating such a slowly
converging quantity.

In this section we use a different techniques to follow the
development of the steady-state channel network via finite size
simulations. At the end we will discuss the subtleties associated with
the slow convergence and the appearance of a second length scale.

%%%%%%%%%%%%%%%%%%%%%%%
\subsection{Histograms}
\label{sec:histo}

In the predicted channel network picture the flow is strictly confined
to certain favorable channels with other sites containing trapped
particles.  

For large systems, the convergence to a fixed channel network means
that lattice outlets can be divided into two types depending on
whether or not they are on the channel network. Once a steady-state
current pattern has been setup, network outlets pass one particle at
every time step and off-network outlets never pass particles. In
finite systems the network is less well defined and the division is
not perfect, but we can see the effect developing by recording for
what fraction of time steps~($s$) each outlet passes a particle and
then plotting a histogram for the fraction of outlets with each
fractional occupation, $h(s)$.

As the system size gets large $h(s)$ will approach two delta functions
at $s=0$ and $s=1$ with the relative weights of each peak determined
by the current flowing,
\begin{equation}
  h(s, J, L\rightarrow\infty)=
  (1-\frac{J}{2})\delta(s)+\frac{J}{2}\delta(s-1).
\end{equation}
These peaks are inconvenient to plot so we instead consider the
integral of $h(s)$. For an infinite system the integral approaches the
constant $1-J/2$ for $0<s<1$. We remove this constant by scaling the
difference between the integral and 1 by $2/J$ and define $H(s)$ by
\begin{equation}
1-H(s)=\frac{2}{J}\left[1-\int_0^s h(\tilde{s})\,d\tilde{s}\right],
\end{equation}
so that $H(s=1)=1$, 
\begin{equation}
\int_0^1H(s)\:ds=0  
\end{equation}
and $H$ should be equal to zero for $0<s<1$ in the infinite size
limit.

%%%%%%%%%%%%%%%%%%%%%%%%
\subsection{Network scaling regime}
\label{sec:scaling}

The histogram function~$H(s,J,L)$ for large~$L$ and small~$J$ has a
scaling form which is very different from that of~$J$,~$B$
and~$\Pi$. The correlation length~$\xi_{n}$, which is the the
characteristic vertical scale of the network---roughly the distance
between nodes, plays an important role. To emphasize this length (and
avoid complicating the situation by introducing~$\xi_{f}$ which
controls the equilibration of the current density) we initially
restrict our attention to systems that have no traps so that the
current is fixed by the initial conditions. The fact that~$J$ is now
an independent variable which we can control is also very helpful.

\begin{figure}
  \begin{center}
    \leavevmode
    \epsfxsize=4truein
    \epsfbox{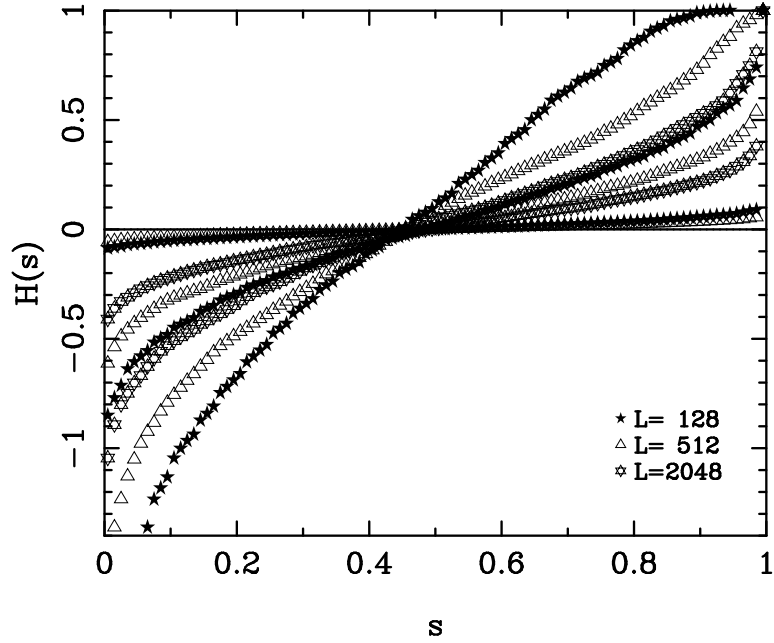}
  \end{center}
  \caption{Basic cumulative histogram
  data from one-deep model using $J=1/16$, $1/4$ and $1$ with system
  sizes $L=128$, $512$, $2048$. For a given $L$, the curves for
  larger $J$ are flatter.}
  \label{fig:hist-raw}
\end{figure}

A series of simulations were carried out with no traps and initially
no more than one particle per site.  Wide systems of width
$W=16L^{1/2}$ running for $3L$~time steps were used with histogram
data collected for the last $L$ time steps. Each simulation was
repeated $N_s$ times with $N_s$ chosen so that $WN_s=5120$ in order to
collect an equal amount of data for each size.  Basic data for
selected $J$ and $L$ values is shown in Fig.~\ref{fig:hist-raw}.

\begin{figure}
  \begin{center}
    \leavevmode
    \epsfxsize=4truein
    \epsfbox{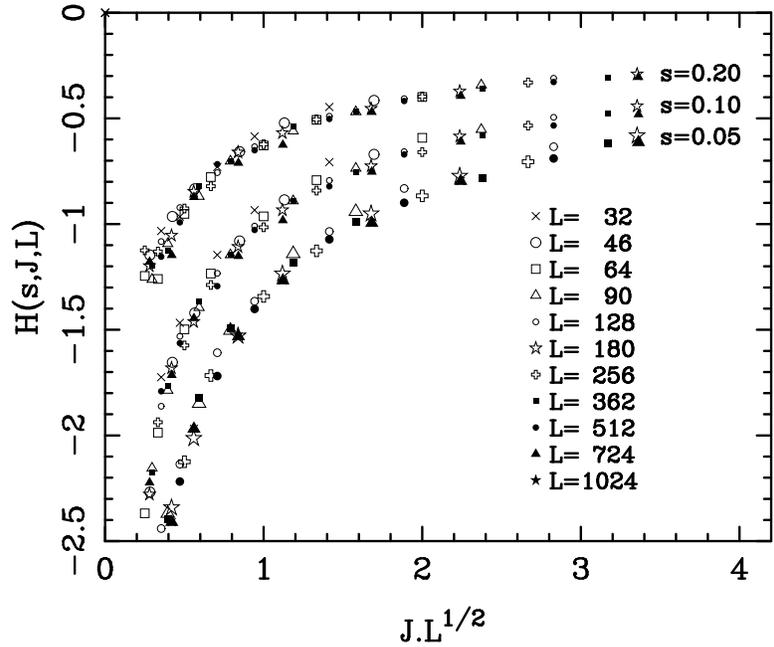}
  \end{center}
  \caption{Scaling of histograms for fixed
  occupancy fraction ($s$) with system size ($L$) and current
  ($J$)}
  \label{fig:fscale}
\end{figure}

With this scaling of the width, the analysis of
Ref.~\onlinecite{watson:disc} %
suggests the finite size scaling hypothesis
in the $W\gg L^{1/2}$~limit,
\begin{equation}
  H(s,J,L)=\tilde{H}_n\left(s,\frac{L}{\xi_{n}(J)}\right).
  \label{eq:hn}
\end{equation}
We expect $\xi_{n}\sim1/J^2$ for small $J$ so if we plot $H(s)$
against $JL^{1/2}$ for different values of $L$ and $J$ at a single
value of $s$ then all the points should fall on a single curve.
Plots for three values of $s$ are shown in Fig.~\ref{fig:fscale}. Data
collapse is quite good with deviations appearing only for the largest
$J$ values. This shows that the relevant correlation length here
appears indeed to be $\xi_{n}$ rather than the correlation length seen
in finite size scaling in Sec.~\ref{sec:crit}, $\xi_f\sim
J^{-\nu/\beta}$. We postpone discussion of the role of $\xi_f$ until
later.

% jplot d-jl-005.plt d-L.dat d-jl-010.plt d-L.dat d-jl-020.plt d-L.dat

\begin{figure}
  \begin{center}
    \leavevmode
    \epsfxsize=4truein
    \epsfbox{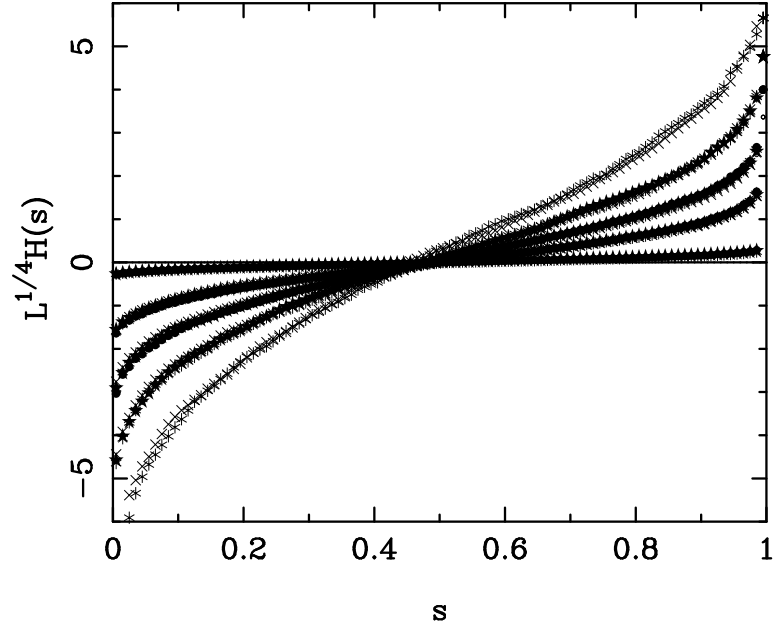}
  \end{center}
  \caption{Length scaled histogram data
  showing overlap for different values of $J$ for the large~$L$ limit.
  Currents used are $J=1$, $J=1/2$, $J=1/4$, $J=1/8$ (only for
  $L>256$) and $J=1/16$ (only for $L>512$)}
  \label{fig:hist-L}
\end{figure}

In addition to data collapse for individual values of $s$ we can also
consider $H$ as a function of $s$ and, with appropriate scaling,
collapse all the data onto a single curve (for a limited range of $J$
and $L$). For large systems, $L\gg\xi_{n}(J)$, we can expand
$\tilde{H}_n$ in $\xi_{n}/L$. The fraction of time each site spends
with the ``wrong'' occupation should decay as $L^{-1/4}$ (from
Ref.~\onlinecite{watson:disc}) so we expect
\begin{equation}
  H(s,L\gg\xi_{n}(J))\approx 
  \tilde{H}_n^{1}(s)\left(\frac{\xi_{n}}{L}\right)^{1/4}.
\end{equation}
since in the limit $L\rightarrow\infty$ we expect $H(s)=0$.  This
means that $L^{1/4}H(s,J,L)$ should collapse onto a set of curves for
different $J$ as $L\rightarrow\infty$. This works rather well as seen
in Fig.~\ref{fig:hist-L} which shows scaled results for 16 different
pairs of $J$ and $L$ values which overlap onto five different curves
for the five different values of $J$. Overlap is least effective for
the smallest $J$ value which corresponds to the largest $\xi_{n}(J)$
where the limit $L\gg\xi_{n}$ is not reached.

\begin{figure}
  \begin{center}
    \leavevmode
    \epsfxsize=4truein
    \epsfbox{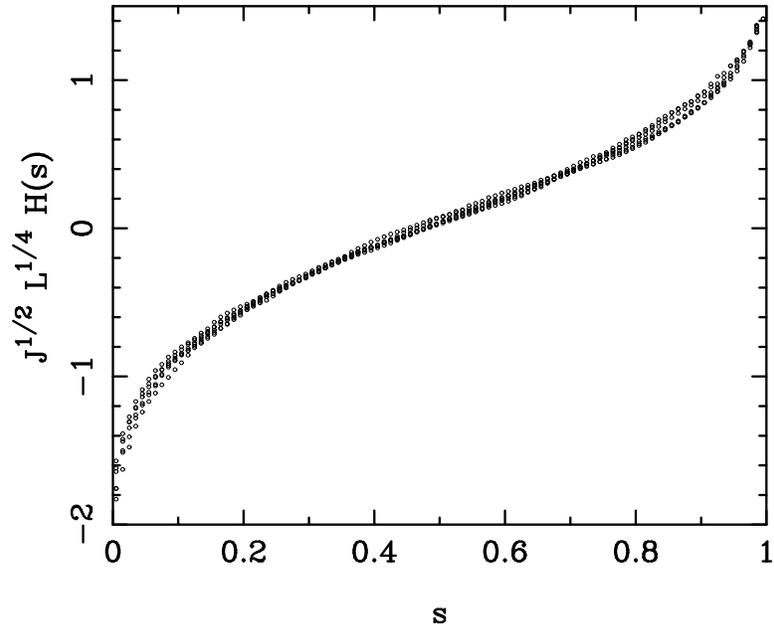}
  \end{center}
  \caption{Data
  collapse for selected parameters: $J=1/8$ for $L=256,512,1024,2048$
  and $J=1/16$ for $L=1024,2048$.}
  \label{fig:hist-collapse}
\end{figure}

Finally we also know that $\xi_{n}(J)\sim J^{-2}$ for small $J$, so
$H(s,J,L)$ should be the same for the same values of $J^2L$ for
$J\ll1$. This works quite well, and when combined with the $L$-scaling
we expect it to produce collapse onto a single curve in a limited
regime:
\begin{equation}
  (J^2L)^{1/4} H(s,J,L)\approx\tilde{H}_n(s) \:\:\:\:\:\:\:\:\:
  \mbox{if $J\ll1$ and $J^2L\gg1$}
\end{equation}
The collapse is quite good, a plot is shown in
Fig.~\ref{fig:hist-collapse}, although the regime of applicability is
rather small as we also need the number of particles in the system
$\sim JL^{3/2}$ to be not too small in order to have sufficient data.

%%%%%%%%%%%%%%%%%%%%%%%
\subsection{Critical scaling regime}
\label{sec:critscale}

So far, we have only investigated the local current distribution in
the regime where $L\sim\xi_n\sim1/J^2$ or larger, i.e., the regime
important for formation of the steady-state network. But from the data
of Fig.~\ref{fig:logj} we can see that the mean current density has
converged to its large system---equivalent to long time---limit when
$L\approx\xi_f\ll\xi_n$. What then is the distribution of local
current densities in this critical scaling regime?

A simple scaling argument suggests the answer: In a wide system of
length $L\sim\xi_f$ (with periodic boundary conditions), each section
of width $\xi_{f\perp}\sim\xi_f^\alpha$ will have of order one (but
sometimes zero as can be seen from the behavior of $\Pi$) current paths
in the steady-state. Thus a fraction of order $1/\xi_f^\alpha$ of the
sites will carry a steady-state current. Since the total current
density is $J\sim\xi_f^{-\beta/\nu}$, the typical (time averaged)
current through theses sites will be
\begin{equation}
  s\sim J^{1-\alpha\nu/\beta}.
\end{equation}
We can thus guess a scaling form for the scaled cumulative
distribution~$H(s)$ in this regime:
\begin{equation}
  H(s,J,L)\approx\frac{1}{J^{1-\alpha\nu/\beta}}
    \tilde{H}_f\left(
    \frac{s}{J^{1-\alpha\nu/\beta}},
      LJ^{\nu/\beta}\right)
      \label{eq:hf}
\end{equation}
where for simplicity we have again simplified to the large~$W$~($\gg
L^\alpha$) limit. For convenience we have used~$J$ instead of $p-p_c$
as the scaling variable, although, of course now $J$~is determined by
the dynamics---including collisions and trap-filling---on length
scales $\lesssim\xi_f$.

%%%%%%%%%%%%%%%%%%%%%%%
\subsection{Intermediate regime and scaling form}
\label{sec:intermed}

For consistency we expect that for {\em large\/} values of the scaling
arguments~$\tilde{H}_f$ should match with the network scaling
function~$\tilde{H}_n$ [Eq.~(\ref{eq:hn})] in the limit of {\em
  small\/} values of the latter's scaling arguments. Thus for 
\begin{equation}
  \xi_f\ll L\ll \xi_n
\end{equation}
we expect a {\em single\/} argument scaling function whose argument
must be a product of the scaling arguments in {\em both\/}
regimes. This yields a unique choice of the argument:
\begin{equation}
  H(s,J,L)\approx\frac{1}{s}\,\tilde{H}_x\left(
  \frac{s}{(JL^2)^\epsilon}
  \right)
  \label{eq:hx}
\end{equation}
with
\begin{equation}
  \epsilon=\frac{2(\beta-\alpha\nu)}{2\beta-\nu}.
\end{equation}
%This can be explicitly tested by running two simulations at the same
%current density, one without traps as above, but with $LJ^2\ll1$, and
%the other with traps, but at $L\gg\xi_f$. Such a comparison is shown
%in Fig.~\ref{fig:makeme}.
% Comments?
% Difficulities with range of L?

In the intermediate regime, {\em neither\/} collisions {\em nor}
critical effects should play much role. Thus we should be able to
understand the scaling of Eq.~(\ref{eq:hx}) from simple considerations.
Regions of length~$\ell\gg\xi_f$ will contribute current to the
drainage basin and hence to the steady-state current unless they happen
to be the exponentially rare regions which sit in anomalously large
holes in the drainage basin. Furthermore, regions much further apart
than~$\xi_f$ (or $\xi_{f\perp}$) will contribute roughly independent
currents. Thus on scales~$\gg\xi_f$, the current will collect on
drainage trees with contributions from each region of length
$\ell\gg\xi_f$ being simply
$J$~times its area (with small variations around this). As soon as one
particle can pass down a drainage tree, then this branch must have no
traps so that traps cannot play a role in the behavior of large
drainage trees. Also, by assumption---to be verified
later---collisions are not important in this regime. The {\em
  statistics\/} of the drainage trees, the current network, and the
local current density on them in this intermediate regime can thus
{\em only depend\/} on~$J$, $L$, and properties of the {\em primary
  outlet trees;\/} they are therefore entirely determined by {\em
  random walk statistics}. The expected scaling for local current
densities is then very simple: the periodic boundary conditions from
bottom to top imply that the current network consists of the primary
outlet paths from all points in the top row which emerge at the {\em
  same\/} point in the bottom row. These will be separated by
horizontal distances of order~$L^{1/2}$ and thus have basins of
width~$L^{1/2}$ implying time averaged local currents on the network
of typically
\begin{equation}
  s\sim JL^{1/2}
\end{equation}
with a distribution determined by the random walk properties. This is
consistent with Eq.~(\ref{eq:hx}) only if $\epsilon=1$. This implies
that $\alpha=1/2$ exactly. Indeed from the above discussion, we could
have guessed this from the observation that the system in this regime
does not ``care'' about the
physics%
\cite{footnote:match,footnote:alpha} %
that determines~$\beta/\nu$.

To check that the above argument is consistent, we need only to check
that collisions are rare in this intermediate regime: since
\begin{equation}
  s\sim JL^{1/2}\ll1,
\end{equation}
this immediately follows.

\begin{figure}
  \begin{center}
    \leavevmode
    \epsfxsize=4truein
    \epsfbox{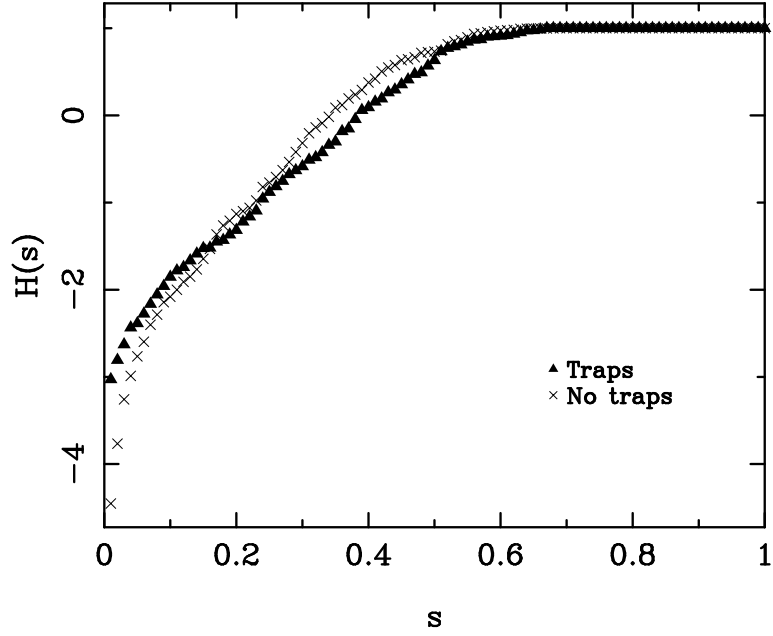}
  \end{center}
  \caption{Histograms for current distribution
  from initial conditions with and without traps. Using a system of
  size 512x360. System with traps has p=0.35 and an average current of
  $J=0.0134$. Current without traps is chosen to be $J=0.0134$.}
  \label{fig:d2a}
\end{figure}

\begin{figure}
  \begin{center}
    \leavevmode
    \epsfxsize=4truein
    \epsfbox{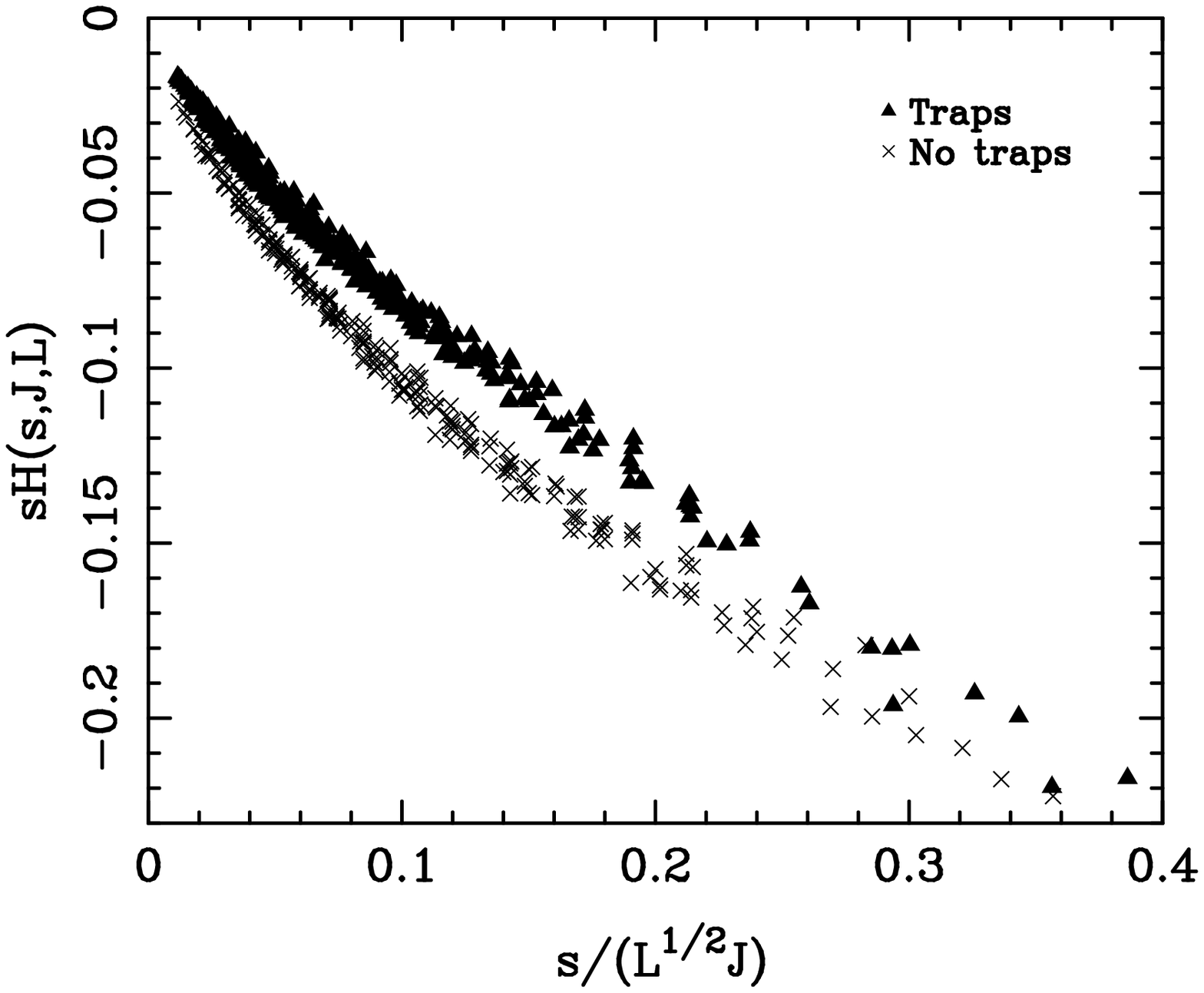}
  \end{center}
  \caption{Test of the scaling form for $H(s)$ in
  the intermediate regime [Eq.~(\protect{\ref{eq:hx}})], using a range
  of system sizes from $L=128$ to $L=1024$ with $p$ and $J$ chosen so
  that $L^2\approx\xi_f\xi_n$ and with $0.02\leqslant
  s\leqslant0.10$.}
  \label{fig:de}
\end{figure}

Results from simulations that try to reach this intermediate regime
are shown in Figs.~\ref{fig:d2a}~and~\ref{fig:de}. Initial conditions
that include traps (fixed~$p$) and exclude traps (fixed~$J$) are
compared for similar mean $J$. Fig.~\ref{fig:d2a} shows $H(s)$ for the two cases averaged
over 28~systems of a single size. The parameters chosen correspond to
$J^2L\approx0.1$ and $L/\xi_f\approx2.5$ [estimating
$\xi_f\approx(p-p_c)^{-\nu}$ with $J\approx2(p-p_c)^\beta$ from
Fig.~\ref{fig:logj}]. To be in the intermediate regime we need
$J^2L\ll1$ and $L\gg\xi_f$; if we neglect the difference between $\nu$
and $\beta$ then this means we need
\begin{equation}
  J^2L \ll 1 \ll JL.
\end{equation}
This is difficult to achieve because the minimum accessible current
density is limited by the size of the system. Despite this difficulty,
the data collapses to the scaling form of Eq.~(\ref{eq:hx}) quite well
(Fig.~\ref{fig:de}). However, the two different initial conditions do
not appear to have converged to the same scaling function in the
accessible range of the parameters.  Nevertheless, our conjectures on
the intermediate regime would seem to be consistent with the numerics.

The existence of an intermediate regime with $L\gg\xi_f$ but
collisions still unimportant, is closely linked to the arguments of
Sec.~\ref{sec:equiv} that collisions are unimportant at long times near
threshold. To see this, consider a system just below threshold
with~$\delta=F_c-F$  and correlation length $\xi_f\sim\delta^{-\nu}$,
which is the length of the longest typical drainage tree. If the force
is suddenly increased by, say,~$\delta/2$, a number of particles of
order~$\delta\xi_f^{d_f}$ will overflow on a tree of length $\xi_f$
resulting in an average number of particles appearing at the same time
at the bottom of the tree of order
\begin{equation}
  n_s\sim\delta\xi_f^{d_f-1}.
\end{equation}
(They will arrive over a time interval~$\xi_f$.) The natural guess is
that the condition for collisions to be irrelevant at the critical
point is that for~$\xi_f$ large, we must have~$n_s\ll1$, i.e., that
\begin{equation}
  1-\nu(d_f-1)>0.
\end{equation}
Using the equalities of Sec.\ref{sec:equiv}, $d_f=1+\alpha-\Gamma/\nu$,
$\alpha=1/2$ and $\beta=1+\Gamma$, this is equivalent to
\begin{equation}
  \frac{\beta}{\nu}>\frac{1}{2}
  \label{eq:bnuineq}
\end{equation}
which is identical to the condition that $\xi_n\gg\xi_f$ just above
threshold. We thus conclude that because of the inequality
[Eq~(\ref{eq:bnuineq})] collisions are dangerously irrelevant at the
critical point---but crucial for the flowing phase at long
times; that adiabatic and sudden forcing are in the same universality
class; and that all the scaling laws should be correct leaving just
two independent exponents ($\beta$ and~$\nu$).  In three dimensions,
where mean field theory results should hold up to
logarithms%
,\cite{narayan:nlflow} %
one similarly finds that collisions should be dangerously irrelevant
near but above threshold.

%%%%%%%%%%%%%%%%%%%%%%%%%%%%%%%%%%%%%%%%%%%%%%%%%%%%%%%%%%%%%%%%%%%
\section{Applications and Conclusions}
\label{sec:appconc}

This paper has investigated a new universality class of dynamic
critical phenomena, which appears to be rather robust. Perhaps the
most interesting aspect of the scaling is the presence of two lengths
above threshold, or, equivalently, two different time scales for
equilibration of the current ($\xi_f$) and of the flow pattern
($\xi_n$). We showed how the histograms of current distributions
indicated the presence of the longer length ($\xi_n$) and how to
reconcile this scaling (with $\xi_n$) with that of $\xi_f$. The other
surprising result is that the critical behavior seems to be in the same
universality class regardless of whether the force increase is sudden
or adiabatic in spite of the dependence of the critical force on the
history. This observation and analytic arguments suggest that dynamic
particle collisions are {\em dangerously irrelevant\/}: they do not
effect the critical exponents unless they are completely absent in
which case the behavior is very different with all of the particles
becoming concentrated onto a single large tree at very long times.

The model we have studied updates all of the sites synchronously so
that particles on different rows can never interact or intermingle. A
more realistic model would allow some inter-row diffusion. The fact
that sudden and adiabatic forcing seem to give the same critical
behavior suggests that the exponents are also likely to be independent
of the order in which the particles are moved and other
time-independent changes in the local rules such as allowing for
position or occupation dependent local particle velocities. However
allowing inter-row diffusion {\em will\/} change the long time
approach to the network (on times greather than~$\xi_n$) from
$t^{-1/4}$ to $t^{-1/2}$ as explained in
Ref.~\onlinecite{watson:disc}.

Previously Narayan and Fisher% 
\cite{narayan:nlflow} % 
studied a related model with a continuous fluid instead of discrete
particles and adiabatic force increase, obtaining numerical results
below threshold and using scaling arguments to infer the properties of
the current network above threshold. They measured
\begin{equation}
  d_f=1.21\pm0.02
\end{equation}
using scaling and assuming that $\alpha\equiv1/2$. This is very close
to our value for discrete particles ($d_{f}=1.22\pm0.05$, the larger
error arising from the uncertainty in $\alpha$, if $\alpha=1/2$ is
assumed exact then our error is also $\pm0.02$). They also obtained
\begin{equation}
  \nu=1.76\pm0.02
\end{equation}
which can be compared to our value of~$\nu=1.62\pm0.04$. The
difference is twice the apparent errors but, as the measurements were
made by different methods, agreement is certainly reasonable. We
conclude that the below threshold behavior (and thus $\xi_f$) of both
continuous and discrete particles are in the same universality class,

But above threshold the models differ because the continuous model has
a different behavior of~$\xi_n$, since the average depth of the rivers
approaches zero at threshold.  In the continuous fluid model, $\xi_n$
depends on the behavior of the probability density,~$p(b)$, that the
secondary outlet barrier is a small amount,~$b$, higher than the
primary barrier. For $p(b)\sim\text{const.}$ as $b\rightarrow0$,
$\xi_n\ll\xi_f$ near threshold. This implies that, in contrast to our
discrete model, the river splits must play a role near threshold if
the force is increased rapidly, a situation not considered in
Ref.~\onlinecite{narayan:nlflow}.

We conclude with a few comments on possible connections between the
critical phenomena discussed here and vortex motion in dirty
superconducting films.

One possible behavior of vortices that is very different from that
found in our model is for the time averaged vortex current in most
regions of space to be non-zero even near to the critical current
threshold. Because of strong tendencies of a vortex lattice to break
up near threshold,
% ref Coppersmith and Millis?
this motion would have to involve sections of vortices moving together
but not synchronously with neighboring regions. It may be possible
that such non-uniform irregular motion with different regions moving at
different times could persist in steady-state.
% Num ref?
But this could also be a transient phenomenon with the system
eventually settling down, near the critical current, to a steady-state
pattern of channels separated by wide regions with
vortices at rest. If this is the case, it is quite possible that, given
the degree of robustness of the critical behavior found here so far,
the critical current phenomena would be in the same universality class
as our model even with the complications of longer range vortex
interactions, particles stopping and starting, history dependence of
the critical current, etc.

A signature of the channel behavior could be found by constructing a
histogram of the time averaged local vortex velocities which should
show a large peak at zero velocity near threshold and some
distribution---with small total weight---at non-zero velocities, in
the simplest scenario centered on a velocity that does not vanish at
threshold. Behavior qualitatively like this has been seen in
simulations of vortices in dirty thin films%
.\cite{faleski:vortdyn} %
Other possible experimental probes of channel flow will be discussed
elsewhere.

It should be noted that the critical behavior found here is
restricted---at least in the model---to the first increase of the
force. Decreasing the force and subsequent increases can result in
different behavior. This and other qualitative phenomena that can
occur in these types of models and vortex systems will also be
discussed elsewhere.

%%%%%%%%%%%%%%%%%%%%%%%%%%%%%%%%%%%%%%%%%%%%%%%%%%%%%%%%%%%%%%%%%%%
\acknowledgements

We thank A.~A. Middleton and S. Redner for useful discussions.  This
work was supported in part by the National Science Foundation via
grants DMR 9106237 and DMR 9630064 and via the Harvard University
Materials Research Science and Engineering Center.

%%%%%%%%%%%%%%%%%%%%%%%%%%%%%%%%%%%%%%%%%%%%%%%%%%%%%%%%%%%%%%%%%%%
%%%%%%%%%%%%%%%%%%%%%%%%%%%%%%%%%%%%%%%%%%%%%%%%%%%%%%%%%%%%%%%%%%%

\newpage

\begin{thebibliography}{10}
\bibitem{footnote:unclear} It is possible that the nonmonotonic
  convergence to scaling observed in the drainage basin may somehow be
  related to the presence of a second characteristic length scale
  above threshold, although why this should appear more strongly here
  than in~$J$ is unclear.
\bibitem{footnote:match} Note that we did not need to use matching to
  the collision dominated network regime to derive the result
  $\alpha=\frac{1}{2}$; in a
  collisionless system, the $L=\infty$ limit is pathological, but the
  physics discussed in the text is only consistent with the large~$L$
  limit
  of Eq.~(\protect{\ref{eq:hf}}) if $\alpha=1/2$.
\bibitem{footnote:alpha} To obtain $\alpha=1/2$, it appears (at this
  point) that one has to consider the physics above threshold. This is
  somewhat analogous to certain conventional dynamic critical
  phenomena, such as superfluid~$^4$He, in which a simple expression
  for the dynamic exponent~$z$ can be obtained readily from physical
  arguments in one phase (the superfluid) but not in the
  other.\protect{\cite{hohenberg:dynrev}}
\end{thebibliography}
\end{document}